\documentclass{emulateapj-rtx4}
\slugcomment{{\sc Accepted to AJ:} May 23, 2012} 
\shorttitle{The Galaxy Luminosity Functions in the Coma Cluster}
\shortauthors{Yamanoi et al.}

\begin{document}


\title{The Galaxy Luminosity Functions down to $M_R=-10$ \\
       in the Coma Cluster\altaffilmark{1}}


\author{
Hitomi Yamanoi\altaffilmark{2},
Yutaka Komiyama\altaffilmark{2,3},
Masafumi Yagi\altaffilmark{2}, 
Sadanori Okamura\altaffilmark{4}, 
Masanori Iye\altaffilmark{2,3},
Nobunari Kashikawa\altaffilmark{2,3}, 
Tadafumi Takata\altaffilmark{3,5},
Hisanori Furusawa\altaffilmark{5},
Michitoshi Yoshida\altaffilmark{6}
}
\email{yamanoi.hitomi@nao.ac.jp}
\altaffiltext{1}{Based on data collected at Subaru Telescope, 
which is operated by the National Astronomical Observatory of Japan}
\altaffiltext{2}{Optical and Infrared Astronomy Division, National Astronomical Observatory of Japan,
2-21-1, Osawa, Mitaka, Tokyo 181-8588, Japan}
\altaffiltext{3}{Department of Astronomical Science, School of Physical Sciences, 
The Graduate University for Advanced Studies (Sokendai), National Astronomical Observatory of Japan,
2-21-1, Osawa, Mitaka, Tokyo 181-8588, Japan}
\altaffiltext{4}{Faculty of Science and Engineering, Hosei University,
3-7-2 Kajino-cho, Koganei, Tokyo 184-8584, Japan}
\altaffiltext{5}{Astronomy Data Center, National Astronomical Observatory of Japan, 
2-21-1, Osawa, Mitaka, Tokyo 181-8588, Japan}
\altaffiltext{6}{Hiroshima Astrophysical Science Center, Hiroshima University,
1-3-1, Kagamiyama, Higashi-Hiroshima, Hiroshima 739-8526, Japan}



\begin{abstract}

We derived the luminosity function (LF) of dwarf galaxies
in the Coma Cluster down to $M_R=-10$ at three fields
located at the center, intermediate, and outskirt.
The LF ($-19<M_R<-10$) shows
no significant differences among the three fields.
It shows a clear dip at $M_R\sim-13$, and is composed
of two distinct components of different slopes; the bright
component with $-19<M_R<-13$ has a flatter slope than
the faint component with $-13<M_R<-10$
which has a steep slope.
The bright component ($-19<M_R<-13$)
consists of mostly red extended galaxies
including few blue galaxies whose colors
are typical of late-type galaxies.
On the other hand, the faint component ($-13<M_R<-10$)
consists of largely PSF-like compact galaxies.
We found that both these compact galaxies and some extended galaxies
are present in the center while only compact galaxies are seen
in the outskirt.
In the faint component, the fraction of blue galaxies is larger
in the outskirt than in the center.
We suggest that the dwarf galaxies in the Coma Cluster,
which make up the two components in the LF, are heterogeneous
with some different origins.

\end{abstract}



\section{INTRODUCTION}

The galaxy luminosity function (LF) is a useful measure to describe
fundamental statistical properties of galaxy populations and serves
as a clue to study the history of galaxy formation and evolution.
Since the monumental work of the LF by \citet{San85},
relatively bright part of the LF has been studied extensively
both for cluster galaxies (e.g., \citealt{Pop05}) and
for field galaxies (e.g., \citealt{Blan03}).
However, our knowledge of the faint part of the LF
($M_R>-19$), which is dominated
by dwarf galaxies, is still very limited.
This is due to the difficulty in determining the distance
of a large number of intrinsically faint galaxies.
Published results of the faint part of the LF, especially,
those of the faintest part ($M_R>-13$) remain scarce.

Clusters of galaxies are a good target to investigate
the faint end of the LF because some statistical methods
can be applied to identify member galaxies.
In this study we chose the Coma cluster
in order to investigate the faint part of the LF, that is,
to study properties of dwarf galaxies and their
possible dependence on different environments in
this cluster.
The Coma Cluster (Abell 1656) at $z=0.023$ (\citealt{SR99})
is the richest cluster of galaxies in the local Universe.
The virial radius ($r_{200}$) of this cluster is
2.0 $h^{-1}$ Mpc, which corresponds to a virial mass
of $1.9 \times 10^{15}h^{-1}M_{\odot}$
as estimated from a weak lensing measurement (\citealt{Kubo07}).
It is relatively easy to take deep images of this
nearby cluster which enables us to reach the faintest
part of the LF.

The LF of the Coma Cluster was studied by many authors
(e.g., \citealt{Ber95, Biv95, Sec97,Tren98,Adami00, AC02,
Beij02, Mob03, Mil07,Adami08, Adami09}).
Several techniques were used to construct the Coma LF.
The statistical background/foreground subtraction is
historically the main one
(e.g., \citealt{Ber95,Tren98,AC02, Beij02,Mil07}).
Others include the spectroscopic redshift technique
(e.g., \citealt{Adami00,Mob03, Adami09}),
photometric redshift technique (e.g., \citealt{Adami08}) and
color magnitude relation technique (e.g., \citealt{Biv95,Sec97}).
The faint-end slope of the LF derived from previous studies
do not agree with each other and it seems
that it depends on the magnitude ranges where the slope
was measured, i.e., the slope is flatter at brighter magnitudes,
and it becomes steeper at fainter magnitudes.
However, a simple comparison of the values of the slope
may be inappropriate because of the difference in the
observed area and galaxy sampling criteria among
previous studies.
Most previous studies observed the cluster core only,
with a few probed outskirts as well as the core.

The aim of the present study is to investigate the behavior of
the faint part, especially, the faintest part of the LF
of Coma cluster using a statistically significant large sample
constructed from deep and wide photometric images obtained with
the Subaru Suprime-Cam.
Our imaging data cover much wider area than most of the previous
studies that reached similar limiting magnitudes.
We employ a careful procedure for subtracting the background
galaxy population statistically
to derive the LFs of the cluster galaxy population.

We investigate the properties of the very faint dwarf
population and their possible environmental dependence, if any,
in terms of LF, color, and surface brightness.
Our galaxy catalogs for the core and the outer areas are constructed
by using the same instrument, same filters, and the
same selection criteria, which ensures coherent and
homogeneous photometric data to allow credible
comparisons of galaxy properties in different environments
of the cluster.

We adopt the standard cosmology model with
$H_0=70$ km s$^{-1}$ Mpc$^{-1}$, $\Omega_m=0.3$, and
$\Omega_\Lambda=0.7$. Magnitudes are on the AB system.
We assume a distance to the Coma Cluster of 100 Mpc
and a distance modulus of 35.0
throughout this paper.

%%%%%%%%%%%%%%%%%%%%%%%%%%%%%%%%%%%%%%%%%%%%%
%%%%%%%%%%%%%%%%%%%%%%%%%%%%%%%%%%%%%%%%%%%%%
%%%%%%%%%%%%%%%%%%%%%%%%%%%%%%%%%%%%%%%%%%%%%
%%%%%%%%%%%%%%%%%%%%%%%%%%%%%%%%%%%%%%%%%%%%%
%%%%%%%%%%%%%%%%%%%%%%%%%%%%%%%%%%%%%%%%%%%%%

\section{OBSERVATIONS AND DATA}

%%%%%%%%%%%%%%%
\subsection{Imaging Data}

\begin{figure}
\centering
\includegraphics[width=6cm]{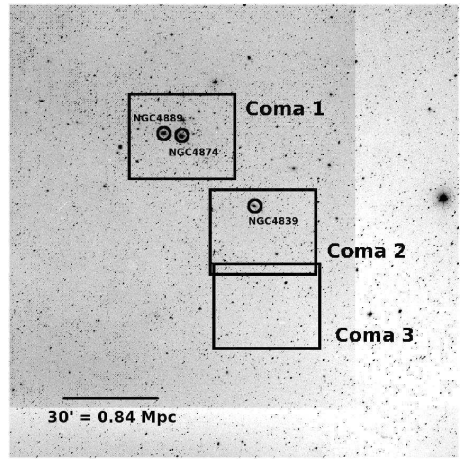}
\caption{
Digitized Sky Survey (DSS) (1st) image of the Coma Cluster. 
North is up, and East is to the left.
The rectangles show the Coma 1 (center), Coma 2 (NGC 4839 groups), and Coma 3 (outskirt) fields.
The three giant elliptical galaxies are marked with circles,
NGC 4889, NGC 4874 and NGC 4839.
\label{fig:coma_field}
}
\end{figure}

\begin{figure}
\centering
\includegraphics[width=6cm]{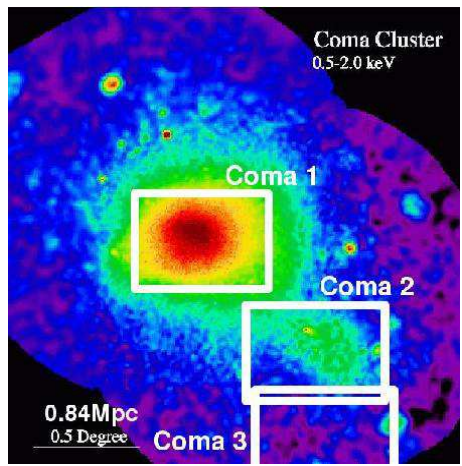}
\caption{
ROSAT X-ray image of the intracluster gas in the Coma Cluster from NASA HEASARC.  
North is up, and East is to the left. 
The color scale shows higher X-ray intensity for red and lower one for blue. 
\label{fig:coma_x}
}
\end{figure}

We perform $B$- and $R$-band observations with Suprime-Cam (\citealt{Miya02})
mounted on the Subaru Telescope  (\citealt{Iye04}) for a total of three nights,
between 12 and 14 May 2007.
Observing conditions were mediocre due to the occasional cloud traversals
on all three nights.
The Suprime-Cam covers a $34' \times 27'$ field of view with a pixel
scale of $0''.202$ pixel$^{-1}$.
A typical seeing size was $\sim 0.''8$ in $B$- and $R$-band.
We selected three fields in the Coma Cluster for the present study, and
they are hereafter referred to as the
Coma 1, Coma 2, and Coma 3 fields, respectively.
The Coma 1 field is located in the central region of the Coma Cluster.
Coma 2 is located at 1.3 Mpc projected distance to the southwest of the cluster core,
in the postulated infalling region (\citealt{Biv96, Neu01}).
Coma 3 is about 1.8 Mpc away from the cluster core, i.e., the outskirts.
Figure \ref{fig:coma_field} shows the three observed fields on the digitized sky survey (DSS) image.
The distribution of X-ray emission  is shown
in Figure \ref{fig:coma_x}.
We summarize the details of observation
of the Coma Cluster in Table \ref{tab:Coma_data}.

%%%%Table 1
%\clearpage
%\vspace{5cm}
\begin{table*}
\begin{center}
{\scriptsize
\begin{tabular}{crcccccc}
\hline
Region & \multicolumn{1}{c}{Date}    &                    R.A.                    &    Dec.              &  Band   &
Exp. Times &  Lim. Mag. & Effective Area\\
       &              &                  (J2000.0)                 &     (J2000.0)        &         &
(min)  & (5$\sigma$) & (arcmin$^2)$\\ 
\hline
Coma 1 & 12 Mar. 2007 & $12^\mathrm{h}59^\mathrm{m}46^\mathrm{s}$  & $+27^\circ54'02\,''$  &  $B$   & $60$  & 26.0 & 769\\
       & 13-14 Mar. 2007  &                  &                    & $R$  & $35$  & 25.9 &   \\
Coma 2 & 12 Mar. 2007 & $12^\mathrm{h}57^\mathrm{m}22^\mathrm{s}$  & $+27^\circ20'02\,''$  &  $B$   & $60$  & 26.5 & 880\\
       & 12-13 Mar. 2007  &                  &                    & $R$  & $90$  & 25.9 &   \\
Coma 3 & 12 Mar. 2007 & $12^\mathrm{h}57^\mathrm{m}23^\mathrm{s}$  & $+26^\circ53'09\,''$  &  $B$   & $60$  & 26.3 & 863\\
       & 12-14 Mar. 2007  &                  &                    & $R$  & $114$  & 26.3 &  \\  
\hline
\end{tabular}
}
\caption[]{
List of the observations for the Coma Cluster.
}
\label{tab:Coma_data}
\end{center}
\end{table*}

%%%%figure 1
%\clearpage
%%%%

%%%%figure 2
%\clearpage
%%%%

%%%%%%%%%%%%
\subsection {Control Field}

In order to obtain the intrinsic LF of cluster galaxies,
we must correct for the effect of contaminant galaxies,
which are mainly located behind
the cluster. To estimate the number of contaminant foreground
and background galaxies, we use the
Subaru Deep Field (SDF; \citealt{Kashi04})
\footnote{
http://soaps.nao.ac.jp/sdf/Project.html
}
as the control field, which is located near the North
Galactic Pole.
The SDF was carefully chosen to avoid bright stars, bright galaxies,
and nearby clusters. This field was observed with Suprime-Cam
and multi-band data based on the common seeing size
of $\sim 0.''98$ are available.
We use the $B$- and $R$-band data taken from the SDF archive.
Their filter sets are identical to those we used for the
Coma Cluster.
SDF data reach much deeper limiting magnitudes than the Coma data.

We consider that the population of Galactic stars in the SDF
is similar to that in the Coma Cluster because the two fields
have almost the same Galactic latitude ($b\sim88^\circ$
for Coma, $b\sim83^\circ$ for SDF).
Accordingly, we use the SDF data to correct for
the effect of Galactic stars in $R>22$ where we do not
apply star/galaxy separation.

We also use the imaging data of the Subaru XMM-Newton
Deep Survey fields (SXDS; \citealt{Furu08})
\footnote{ 
http://www.naoj.org/Science/SubaruProject/SDS/datapolicyJ.html
}
to examine the
effect of the cosmic variance
(see \S \ref{sec: cosmic_variance} for details).
The SXDS fields are also observed with Suprime-Cam and
the same photometric calibrations as the SDF was applied.

%%%%%%%%%%%%%%%%%%%%%%%%%%%%%%%%%%%%%%%%%%%%%
%%%%%%%%%%%%%%%%%%%%%%%%%%%%%%%%%%%%%%%%%%%%%
%%%%%%%%%%%%%%%%%%%%%%%%%%%%%%%%%%%%%%%%%%%%%
%%%%%%%%%%%%%%%%%%%%%%%%%%%%%%%%%%%%%%%%%%%%%
%%%%%%%%%%%%%%%%%%%%%%%%%%%%%%%%%%%%%%%%%%%%%

\section{DATA REDUCTION AND CATALOG}
\label{sec:reduction}

%%%%%%%%%%%%
\subsection{Data Reduction}

Data reduction was carried out in a standard procedure using the
reduction software
for Suprime-Cam; SDFRED (\citealt{YK02a, Ouc04}) and Image
Reduction and Analysis Facility (IRAF).
We perform the point spread function (PSF) equalization by
convolving gaussians
to the reduced Coma images in order to match their PSF size
(FWHM) to that of the SDF image.
The resulting PSF size is 4.9 pixels ($0.''98$) in all the images,
which correspond to $\sim0.45$ kpc at the distance of the Coma Cluster.

%%%%%%%%%%%%
\subsection{Photometric Calibration}

\begin{figure}
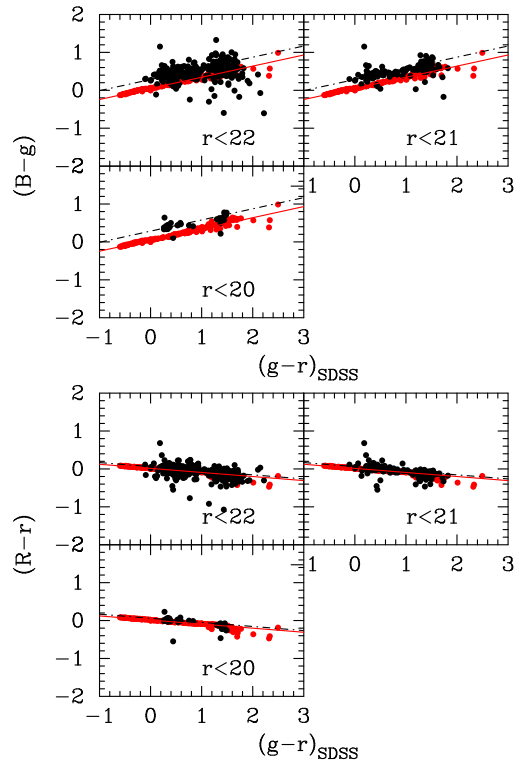

\centering
\includegraphics[width=5cm, angle=270]{f3_Bg.ps}
\includegraphics[width=5cm, angle=270]{f3_Rr.ps}
\caption{
Color-color plots for stellar objects in the Coma 1 field 
with the initial calibration (\textit{black dots}). 
The dot-dashed lines are the least square fit to stars brighter than $r=21$ and $g-r<1.0$
whose $B-g$ and $R-r$ are calculated from our Suprime-Cam measurements
and SDSS catalog.
The colors of Gunn \& Stryker's stars calculated with
the filter models are plotted as red points.
Red lines are the least squares fit to model stars at the range of $g-r<1.0$.
\label{fig:ccd_star_coma1}
}
\end{figure}

First, we carry out an initial estimation of the photometric zero
point using previous photometric 
catalog of \citet{Mob01} and \citet{Komi02}, whose observations
cover our target fields.
However, passbands are slightly different between their catalog
and our data.
In order to improve the accuracy of the zero point, we re-estimate
it using the photometric stars
from the Sloan Digital Sky Survey (SDSS) DR7 photometric catalog
which covers the Coma Cluster and the SDF.
This calibration method is similar to the one by \citet{Yagi10}.
Photometric zero points of all the images are corrected to
give the measured counts (ADU) consistent with the stellar
spectral energy distributions (SEDs)
given in the Bruzual--Persson--Gunn--Stryker Atlas
\footnote{
http://www.stsci.edu/hst/observatory/cdbs/bpgs.html
}
which is an extension of
the Gunn--Stryker Atlas (\citealt{GS83}).
We select the stars that are detected in both the SDSS and the
Suprime-Cam, and
compare the stellar sequences from our data and the Gunn \& Stryker's SED
in the two color planes shown in Figure \ref{fig:ccd_star_coma1}.
Red lines represent the best fits to the Gunn \& Stryker's stars.
Dashed-dotted lines show the
best fit to our raw data for stars brighter than $r=21$,
which are calibrated based on the initial estimation of zero points.
The typical number of stars used for the calibration is about 80 and 30
in the Coma Cluster and the SDF, respectively.
We correct for these offsets in the zero points by shifting
our stellar sequence
to the model sequence computed by convolving 
the Gunn \& Stryker's SEDs with the filter 
response curves.
The offset value is mostly due to the difference of the filter response 
between the Suprime-Cam passbands and 
the filters used in \citet{Mob01} and \citet{Komi02}.
Specifically, we give the offset values; -0.28, -0.36 and -0.23 mag in $B$-band,
-0.07, -0.08 and -0.11 mag in $R$-band of Coma 1, 2 and 3.
The errors in the final photometric zero points are less than 0.02 mag
and 0.03 mag
for the Coma Cluster and the SDF images, respectively.

The Galactic extinction is corrected using the extinction map
by \citet{Schle98}.
Table \ref{tab:zp} lists the photometric zero points and the
Galactic extinction in each band.

%%%%Table 2
%\clearpage
%\vspace{5cm}
\begin{table}
\begin{center}
\begin{tabular}{cccc} 
\hline
Region & Band    & zp      & Galactic Extinction \\
       &         & (mag/ADU) &   (mag)               \\ 
\hline
Coma 1 & $B$ & 33.04 & 0.041\\
       & $R$ & 31.84 & 0.025\\
Coma 2 & $B$ & 33.85 & 0.041\\
       & $R$ & 32.66 & 0.025\\
Coma 3 & $B$ & 34.23 & 0.042\\
       & $R$ & 32.94 & 0.026\\  
SDF    & $B$ & 34.69 & 0.078\\
       & $R$ & 34.26 & 0.048\\         
\hline
\end{tabular}
\caption[]{
List of the photometric zero points and the Galactic extinction 
for the Coma Cluster and the SDF.
}
\label{tab:zp}
\end{center}
\end{table}

%%%%%%figure 3
%\clearpage
%%%% 

%%%%%%%%%%%%
\subsection{Object Extraction and Catalog}
\label{sec:obs_extraction}

We perform object detection and photometric measurement using
the SExtractor (\citealt{BA96})
in double-image mode with the $R$-band image as the detection image.
We identify as an object more than five connected pixels
above $1.5\sigma$ of the sky background noise.
We adopt \verb|MAG_AUTO| for the total magnitudes and $2''$ circular
aperture magnitudes
for colors.

The limiting magnitude is derived by the method similar to that
in \citet{Ouc04}.
We measure the sky counts in a number of $2''$ apertures which
are distributed at random on the image,
and obtain the 1 $\sigma$ noise by fitting a Gaussian function
to the histogram of the sky counts.
We confirm that there is no failure of  a $2''$ aperture
photometry of detected objects
in the  $R$-band
and that measurements of $B$-band aperture magnitude of the
object brighter than
the limiting magnitude are successful in double-image mode.
The $5\sigma$ limiting magnitudes and the effective areas are
summarized in Table \ref{tab:Coma_data}.
The limiting magnitude in each band is found to be $\sim26$
mag ($5\sigma$).
Regions around bright stars and giant galaxies
(for example, NGC 4889, NGC 4874, NGC 4898, NGC 4865,
NGC 4911 and IC 4051 in the Coma 1),
where the detection of faint objects fails, are masked out.

\begin{figure}
\centering
\includegraphics[width=5cm, angle=270]{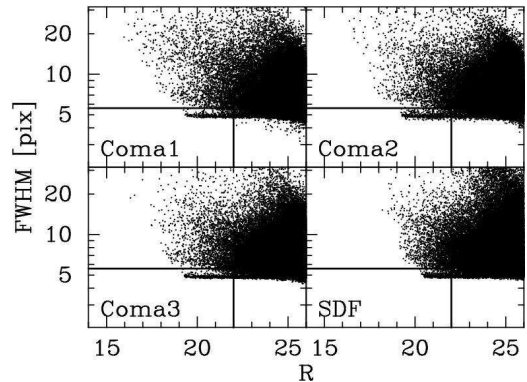}
\caption{
FWHM \textit{vs.} R-magnitude of all objects 
(excluding saturated objects) 
in the three observed fields and the SDF.
The solid line indicates the discrimination between stars and galaxies. 
No star/galaxy discrimination is performed at fainter magnitudes ($R>22$).
\label{fig:Rfwhm}
}
\end{figure}

The star/galaxy discrimination is performed for objects with $R\leq22$.
We regard the objects as galaxies which are significantly
more extended (FWHM$>$5.6 pixels=$1.''13$) than stars
(FWHM$=4.9$ pixels=$0.''98$) as shown in Figure \ref{fig:Rfwhm}.
As for stellar objects fainter than $R=22$,
the statistical subtraction 
(for details in \S \ref{sec:back_sub}) 
is applied in the following analysis.
We confirm that the LFs remains unchanged if we change the star/galaxy
discrimination limits
to $R=21$ or $R=23$ instead of $R=22$.
Details are given in \S \ref{sec:lf}.
We investigate the contamination from globular clusters fainter
than $R=22$ in detail in \S \ref{sec:gc}, and it is found
to be negligible.

\begin{figure}
\centering
\includegraphics[width=5cm, angle=270]{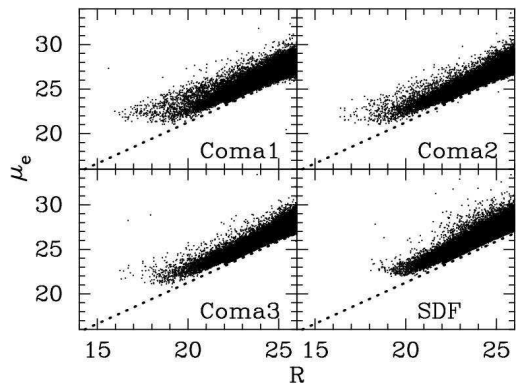}
\caption{
The effective surface brightness \textit{vs.} R-magnitude of objects
in the Coma Cluster and the SDF.
The sequence of PSF-size objects is shown by the dashed line in each panel. 
Note that we exclude saturated objects and stars brighter than $R=22$.
\label{fig:Rmu}
}
\end{figure}

We evaluate the effective surface brightness of the objects
($\mu_e$) in units of mag arcsec$^{-2}$.
It is defined by the mean surface brightness inside the effective radius
(radius in which half the total flux is emitted) and
contains some information on the luminosity profile of the objects.
The distribution of the effective surface brightness as a function of
total $R$-band magnitude is shown in Figure \ref{fig:Rmu}.
The dashed line indicates the sequence of PSF-size objects in each field.
Saturated objects and stars with $R\leq22$ are removed from this figure.

%%%%figure 4
%%%%

%%%%figure 5
%%%%

%%%%%%%%%%%%%
\subsection{Color Classification}
\label{sec: color_classification}

\begin{figure}
\centering
\includegraphics[width=5cm, angle=270]{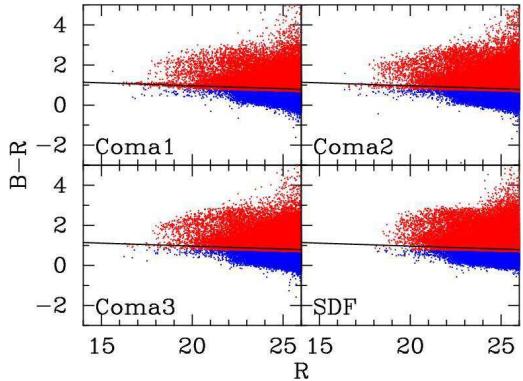}
\caption{
Color magnitude diagram ($B-R$ \textit{vs.} $R$) in the Coma Cluster and the SDF.
The solid line in each panel represents the color magnitude relation of bright galaxies in our Coma 1 region.
We separate red and blue galaxies based on the line shifted blueward by $\Delta(B-R)=-0.2$.
\label{fig:cmd_coma}
}
\end{figure}

We measure the colors of the objects within $2''$ apertures,
which corresponds to $\sim 0.9$ kpc at the Coma Cluster distance.
Figure \ref{fig:cmd_coma} shows
the color-magnitude diagram (CMD) of the objects in the Coma Cluster and SDF.
Note that only galaxies are plotted for $R\leq22$ where star/galaxy discrimination was applied,
while all the objects are plotted for $R>22$.
The color-magnitude relation (CMR), indicated by the solid line in each panel,
is the least-squares fit to all the galaxies in the central region that are spectroscopically
confirmed as members of our Coma 1 field from the \citet{Mob01} catalog.
It is represented by 
\begin{equation}
\label{eq:cmr}
(B-R)=(-0.029\pm0.010)R+(1.55\pm0.21),
\end{equation}
and broadly consistent with the CMR by
$(B-R) = (-0.045\pm0.028)R+(2.27\pm0.48)$ 
in their bandpass of \citet{Adami06a}.

\citet{Adami06b} reported that most of the faint galaxies ($21<R<25$)
follow the CMR derived for brighter elliptical galaxies ($R<18$)
in the Coma Cluster, suggesting that these low-luminosity
galaxies experienced
a passive evolution similar to brighter ones.
\citet{Adami09} confirmed that the spectroscopically identified
dwarf galaxies ($20<R<23$) also follow the CMR. 
Assuming that Equation (\ref{eq:cmr}) is valid down to our limiting magnitude ($R\sim26$),
we divide the galaxies into red and blue populations.
We adopt the separation boundary which is parallel to the
CMR but shifted blueward by $\Delta(B-R)=-0.2$,
i.e, $(B-R)=-0.029R+1.35$.

%%%%figure 6
%\clearpage
%%%%

%%%%%%%%%%%%
\subsection{Completeness and False Detection Estimate}

\begin{figure}
\centering
\includegraphics[width=5cm, angle=270]{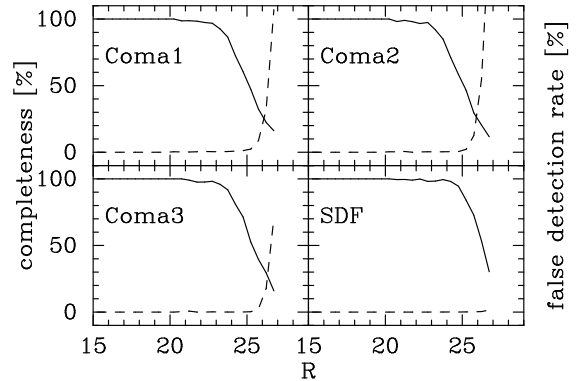}
\caption{
The detection completeness (\textit{solid line}) and the false detection rate (\textit{broken line})
for the Coma Cluster and the SDF as a function of $R$-band magnitude.
\label{fig:comp_false}
}
\end{figure}

Detection completeness of objects in the Suprime-Cam images depends
on the object magnitudes.
We estimate the completeness through a series of simulations by detecting
artificial objects in the original images following the procedure
given in \citet{Kashi04}.
Gaussian profiles with different peaks whose FWHM is equal to
the seeing size are given to artificial objects.
These artificial objects are randomly distributed on the object-subtracted image,
which is -\verb|OBJECTS| check image made by SExtractor.
We mask circular areas around objects with $>10\sigma$ detection threshold
of the sky background noise on the image
to avoid detections of artificial objects embedded above bright objects.
The radius of each circle is FWHM$\times1.5$ of these bright objects.
The artificial star magnitudes are distributed every 0.5 mag
from 20.0 mag to 27.0 mag.
We extract the artificial objects with SExtractor using exactly
the same parameters as in \S \ref{sec:obs_extraction}, and
estimate the completeness factor, $k(R)$, i.e., the number of
the detected objects divided by the number of artificial objects
in the input list as a function of $R$ magnitude.
We generate 250 artificial objects for each magnitude bin and
repeat the process ten times.
The detection completeness for our catalog is shown by the solid curves in
Figure \ref{fig:comp_false}.
Each image has more than 90\% completeness at $R<23.5$.

We also show the false detection rate for $R$-band detected objects
by the broken curves in Figure \ref{fig:comp_false}.
It is estimated by performing photometry on the negative image
at the same threshold level
as the positive i.e. observed image using the SExtractor.
We confirm that the false detection rate is almost $0\%$ down
to $R=25$ in all fields.
Therefore, no correction of the false detection rate is
performed in this study.

We use observed galaxies down to $R=25$ ($M_R=-10$) and perform
the completeness correction by calculating
$N(R)_{\rm corrected}=N(R)_{\rm observed}/k(R)$
for each 0.5 magnitude bin.
The magnitude error of  artificial objects between input and
output  magnitudes is less than 0.05 mag
at $R=25$ in all images.

%%%%figure 7
%clearpage
%%%%

%%%%%%%%%%%%%%%%%%%%%%%%%%%%%%%%%%%%%%%%%%%%%
%%%%%%%%%%%%%%%%%%%%%%%%%%%%%%%%%%%%%%%%%%%%%
%%%%%%%%%%%%%%%%%%%%%%%%%%%%%%%%%%%%%%%%%%%%%
%%%%%%%%%%%%%%%%%%%%%%%%%%%%%%%%%%%%%%%%%%%%%
%%%%%%%%%%%%%%%%%%%%%%%%%%%%%%%%%%%%%%%%%%%%%

\section{CONSTRUCTING LUMINOSITY FUNCTION}
\label{sec:const_lf}

%%%%%%%%%%%%%%%%%
\subsection{Number Density}

\begin{figure}
\centering
\includegraphics[width=6cm, angle=270]{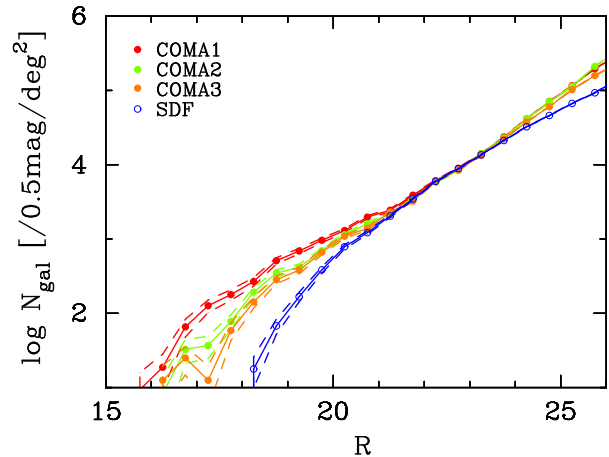}
\caption{
The number density of all the objects in the Coma Cluster and the SDF plotted 
against $R$-band magnitude.
All counts are scaled to an area of 1.0 deg$^2$ and the completeness corrections are performed.
The Poisson statistical errors are shown as dashed lines.  
\label{fig:ncount_all}
}
\end{figure}

\begin{figure}
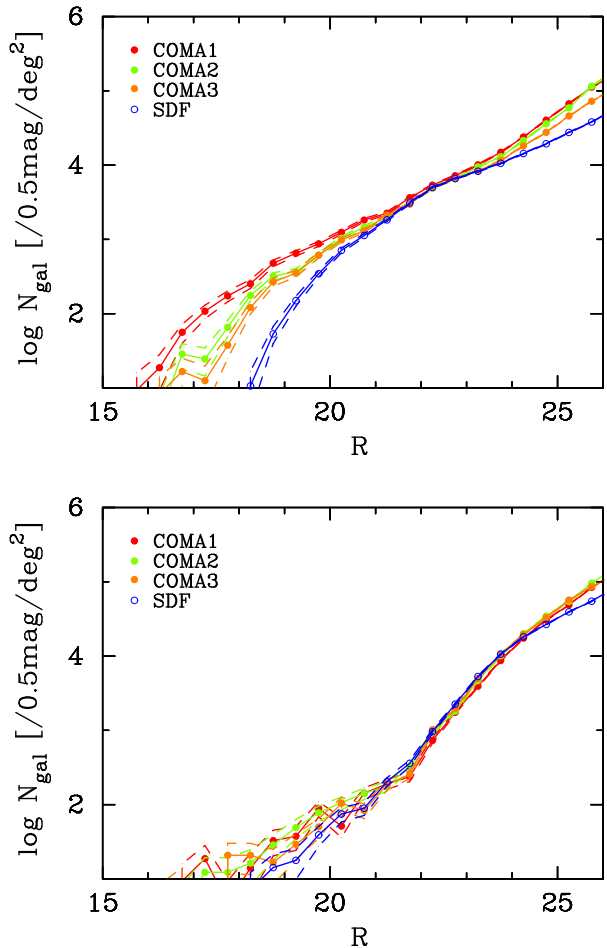

\centering
\includegraphics[width=6cm, angle=270]{f9_top.ps}

\vspace{0.5cm}

\includegraphics[width=6cm, angle=270]{f9_bottom.ps}
\caption{
The number density of red (\textit{top}) and blue (\textit{bottom})
objects in the Coma Cluster and the SDF plotted against $R$-band magnitude.
The color separation is based on the CMR of the Coma 1 region (for details in \S \ref{sec:reduction}).
All counts are scaled to an area of 1.0 deg$^2$ and the completeness corrections are performed.
The Poisson statistical errors are shown as dashed lines.  
\label{fig:ncount_redblue}
}
\end{figure}

First, we derive the number density of all the objects and
red and blue objects as a function of $R$ magnitude
applying the completeness correction.
Figure \ref{fig:ncount_all} shows the number density of all the objects
while Figure \ref{fig:ncount_redblue} shows that of red and blue objects
in the three fields of the Coma Cluster and the SDF.
The error bars show the Poisson errors.

%%%%figure 8
%\clearpage
%%%%

%%%%figure 9
%\clearpage
%%%%

%%%%%%%%%%%%%%%%%%
\subsection{Subtraction of Background/Foreground Galaxies}
\label{sec:back_sub}

The LF is defined as the number density of galaxies per
unit magnitude interval as a function of absolute magnitude:
$dN(M)=\phi(M)dM$, where $dN(M)$ is the number of galaxies
per unit area with magnitudes in the $M$ to $M+dM$ range.
To derive the intrinsic LFs of cluster galaxies,
we must correct for the effect
of contaminant galaxies, which are mainly located behind the cluster.

We use the statistical background subtraction method
(e.g. \citealt{DeP95,Ber95,Phi98,Tren98,Kamb00,YK02b,Pop05,Mil07}).
We perform the subtraction using the surface brightness
versus magnitude diagram shown in Figure \ref{fig:Rmu}. 
We divide the diagram
into meshes with $\Delta\mu_e=1.0$ mag arcsec$^{-2}$ and $\Delta{R}=0.5$ mag
and denote the number of objects in a mesh as $N(\mu_e, R)$.
The number density of member galaxies is given by:
\begin{equation}
\label{eq:estmate_member}
N(\mu_e, R)_{\rm mb}=N(\mu_e, R)_{\rm cl}/k_1(R) -
N(\mu_e, R)_{\rm f}/k_2(R) ,
\end{equation}
where $N(\mu_e, R)_{\rm mb}$ is the number density of member galaxies of
the cluster per unit area (1.0 deg$^{2}$),
$N(\mu_e, R)_{\rm cl}$ is the number density of all the objects obtained
from a cluster field, and $N(\mu_e, R)_{\rm f}$ is the number density of
all the objects obtained from the SDF, the control field.
Here, $k_1(R)$ and $k_2(R)$ are the completeness factors for the
cluster field and control field, respectively (Figure \ref{fig:comp_false}).
This procedure is applied to each of the three Coma fields in
computing the number density of all the objects and red and blue
objects. After the statistical subtraction, we summed
up $N(\mu_e, R)_{\rm mb}$ with respect to $\mu_e$ to obtain the LF
of the cluster $N(R)_{\rm mb}$.

%%%%%%%%%%%%%%
\subsection{The Effect of Cosmic Variance}
\label{sec: cosmic_variance}

In order to check the effect of cosmic variance on the LF,
we re-compute the Coma LF using the SXDS as the alternative
control field.
The SXDS is composed of five sub-fields which are
referred to as Center, North,
South, East and West. Each of the SXDS fields has the same
size of field of view as the SDF  and the Coma fields.
However, the SXDS fields are located at a lower Galactic
latitude than the SDF and the Coma fields
($b\sim88^\circ$ for Coma, $b\sim83^\circ$ for SDF, $b\sim-60^\circ$ for SXDS).

\begin{figure*}
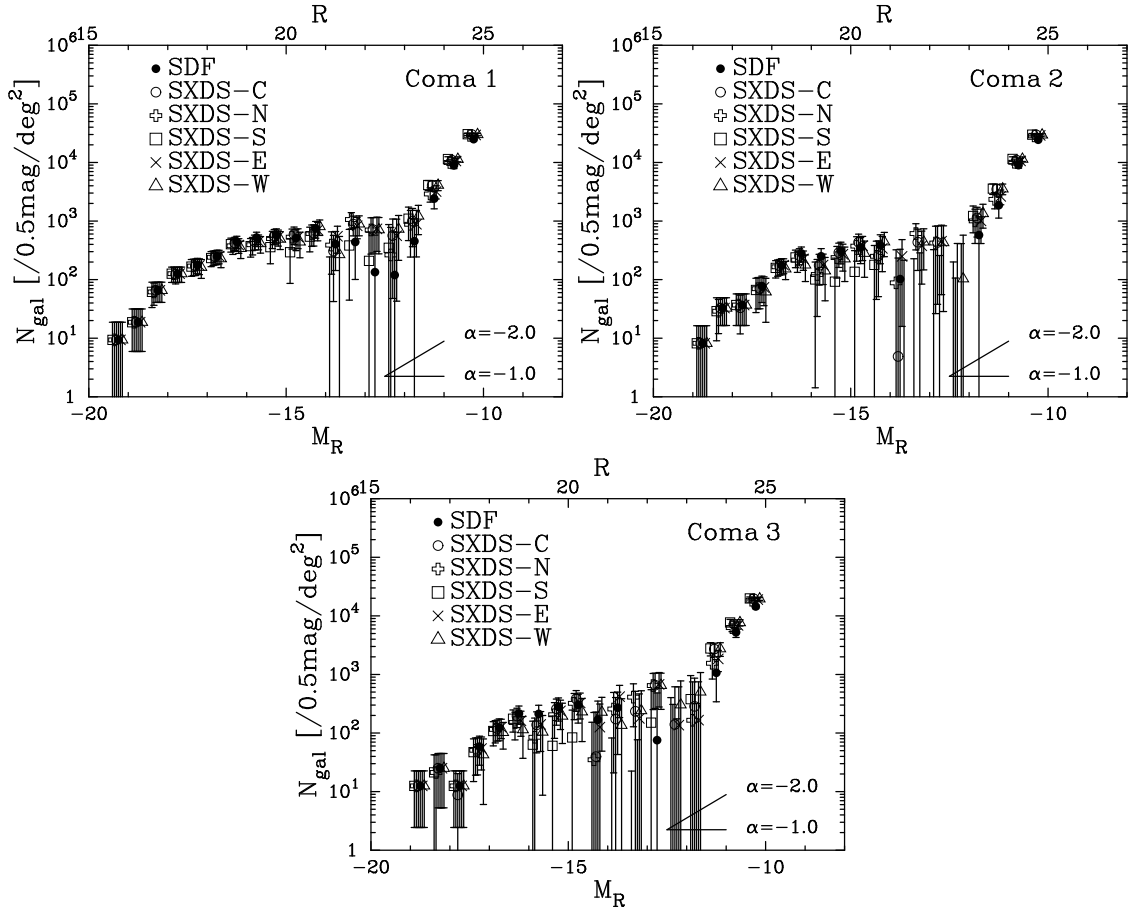

\centering
\includegraphics[width=6cm, angle=270]{f10_coma1.ps}
\includegraphics[width=6cm, angle=270]{f10_coma2.ps}
\includegraphics[width=6cm, angle=270]{f10_coma3.ps}
\caption{
The LFs in the Coma 1 (\textit{top-left}), the Coma 2 (\textit{top-right}) 
and the Coma 3 (\textit{bottom}) obtained by the statistical background subtraction using
different control fields, the SDF and the five SXDS fields.
Note that each of the five SXDS fields is indicated by center (C), north (N), south (S), east (E) and west (W),
respectively.
The error bars of the LFs come from the Poisson statistics.  
\label{fig:Rlf_comp_field}
}
\end{figure*}

We show in
Figure \ref{fig:Rlf_comp_field}
the LFs in the three regions of the Coma Cluster
estimated by the statistical background subtraction using the
the five SXDS fields as the control field.
The errors of each LF are estimated based on Poisson statistics.
Figure \ref{fig:Rlf_comp_field} indicates that the error bars
of all LFs overlap
in the whole magnitude bins in all Coma fields.
When the LFs reach the magnitude of $R\sim-13$, their slopes change.
The two clear components (bright component and faint component)
of different slopes are confirmed in all the LFs.
We conclude that the effect of cosmic variance is insignificant
between the SDF and the SXDS fields.
In the SXDS-South region, some massive structures with masses close
to $10^{14}$ M$_{\bigodot}$ were discovered by \citet{Fino10}.
From their results, the SXDS-North region is not
polluted by background massive structures.
We made the LF shown in Figure \ref{fig:Rlf_comp_field}
without masking the massive structure in the SXDS-South.
We do not see any significant effect of the structure.

In this paper, we adopt the LFs obtained by using the SDF
as the control field because the SDF and the Coma Cluster
are located at nearly the same Galactic latitude,
and the number count of faint stars are similar to
the two fields.

%%%%figure 10
%\clearpage
%%%%

%%%%%%%%%%%%%%
\subsection{Contribution from Globular Clusters}
\label{sec:gc}

At faint magnitudes, we must consider the contamination from globular
clusters (GCs).
Unresolved low-luminosity galaxies whose angular sizes
are similar to the seeing size
cannot be distinguished from bright GCs
that appear as point sources at the distance of the
Coma Cluster.
One of the fundamental parameters of GC study is the
specific frequency (\citealt{HvdB81})
\begin{equation}
\label{eq:Sn}
S_{N}= N_{GC}10^{0.4(M_V+15)},
\end{equation}
where $N_{GC}$ is the total number of GCs in the galaxy.
$S_N$ is normalized to a galaxy with an absolute $V$
magnitude of $-15$.
\citet{M&A02} studied the GCs of 17 elliptical galaxies
in the Coma Cluster and found
a mean value of the specific frequency $S_N=5.1$.
On the other hand, the GC luminosity function (GCLF hereafter)
of  a galaxy is described by
a Gaussian distribution function (\citealt{Harris91}):
\begin{equation}
\label{eq:GCLF}
n_{GC}(m)=\frac{N_{GC}}{\sqrt{2\pi}\sigma}e^{-(m-m^0)^2/2\sigma ^2},
\end{equation}
where $N_{GC}$ is the total number of GCs, $m^0$ is the
turnover (peak) magnitude of the
distribution, and $\sigma$ is the dispersion.
\citet{Harris09}
reports the GCLF parameters for giant elliptical galaxies
in the Coma Cluster as $V^0=27.7$ and $\sigma=1.48$.
We compute the total GCLF around giant red galaxies brighter
than $M_R=-18$ in the Coma 1 field based on Equations (\ref{eq:Sn})
and (\ref{eq:GCLF}).
The GC contamination from galaxies fainter than $M_R=-18$
is much smaller than
from giant galaxies.
No separate estimation of the total GCLFs is performed in  Coma 2
and Coma 3 fields
since the total GCLFs in these fields are supposed to be
below that in the Coma 1 field.

%%%%%%%%%%%%%%%%%%%%%%%%%%%%%%%%%%%%%%%%%%%%%
%%%%%%%%%%%%%%%%%%%%%%%%%%%%%%%%%%%%%%%%%%%%%
%%%%%%%%%%%%%%%%%%%%%%%%%%%%%%%%%%%%%%%%%%%%%
%%%%%%%%%%%%%%%%%%%%%%%%%%%%%%%%%%%%%%%%%%%%%
%%%%%%%%%%%%%%%%%%%%%%%%%%%%%%%%%%%%%%%%%%%%%
\section{RESULTS}
\label{sec:results}

%%%%%%%%%%%%
\subsection{Luminosity Function of All Galaxies}
\label{sec:lf}

\begin{figure}
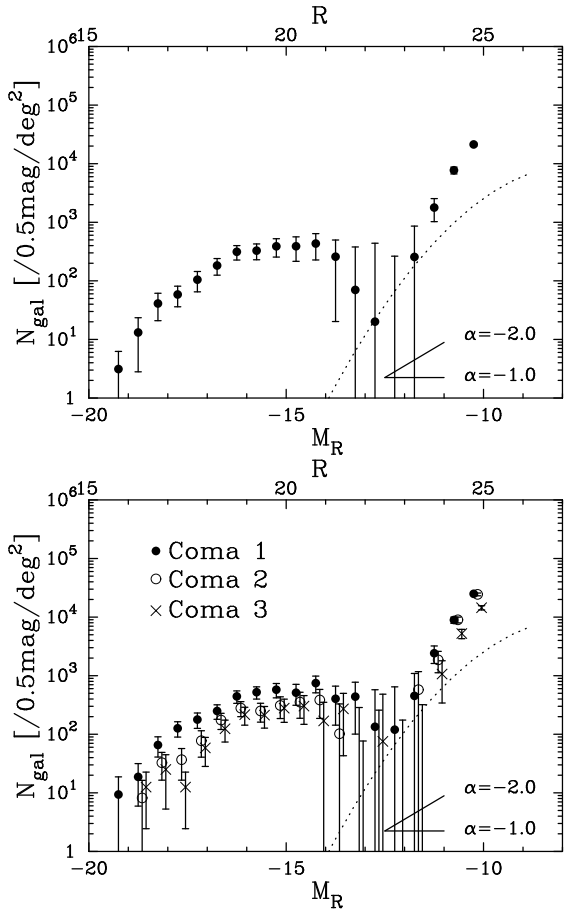

\centering
\includegraphics[width=6cm, angle=270]{f11_top.ps}
\includegraphics[width=6cm, angle=270]{f11_bottom.ps}
\caption{
\textit{Top}: the total $R$-band LF of  Coma 1, Coma 2 and Coma 3. 
\textit{Bottom}: the LFs of
the Coma 1 (\textit{filled circles}), 
Coma 2 (\textit{open circles}) and Coma 3 (\textit{cross marks}) fields, respectively.
The error bars are based on Poisson statistics.
The broken curve indicates the LF of GCs in the Coma 1 field calculated using the previous results 
of \citet{M&A02} and \citet{Harris09}.
All counts are scaled to an area of 1.0 deg$^2$.
Note that the LF plots of Coma 2 and Coma 3 are shifted rightward for the visibility. 
\label{fig:Rlf_all_gc}
}
\end{figure}

The top panel of Figure \ref{fig:Rlf_all_gc} shows the LF of all
the three fields while the bottom panel shows the LFs of
Coma 1, 2, and 3 fields separately.
The broken curve in Figure \ref{fig:Rlf_all_gc} is the
total GCLF of the Coma 1 field computed in \S \ref{sec:gc}.
From Figure \ref{fig:Rlf_all_gc}, we find that the
faintest part of the LF ($M_R>-12$) remains nearly unaffected
by GCs ($<15$\%) because the number of faint
dwarf galaxies at $M_R>-12$ is about 10 times
larger than that of GCs even in the Coma 1.
\cite{Peng11} observed that intracluster GCs in the
Coma Cluster have
a flat spatial distribution within 520 kpc from NGC 4874.
The number count of intracluster GCs at $r=520$ kpc
is less than 1\% of that of GCs
around NGC 4874, and it is considered to be negligible.
Furthermore, we masked the regions around giant
galaxies on our images as mentioned
in \S \ref{sec:obs_extraction}.
Indeed, at least more than 50\% of the total number
of GCs around these galaxies are
excluded from our data, if we refer to \citet{Harris09}.
Therefore, the GCLF shown by the broken curve in
Figure \ref{fig:Rlf_all_gc} can be regarded as the upper limit of the
GC contaminations for both the total LF and for each LF of
the three fields.

The top panel shows the clear signature of two components
comprising the faint part of the LF. One is dominant
at $-19<M_R<-13$ and the other is
dominant at $-13<M_R<-10$. They are hereafter referred to as
the bright dwarf component (BDC) and the faint dwarf component
(FDC), respectively.
These distinct components appear in all the LFs
in spite of the different environments such as the Coma 1, 2 and 3.
No significant difference is found in the 
\textit{shape} of the LF among the three regions.

\citet{Mob03} reported that the bright part of the LF ($-23<M_R<-16$)
is similar in the central and outer regions (including our Coma 2
and Coma 3 fields).
On the other hand, \citet{Adami07a} indicated in their Figure 11
that small-scale spatial variations of the LF exist in the regions
surrounding giant galaxies in the center of the Coma Cluster.
They divided their observed area ($\sim0.5$ deg$^2$) into
20 subregions of ($10' \times 10'$) and examined the LFs of
the objects in the subregions.
In this study we are interested in the possible variation of
the LF over much larger spatial scale than that examined by
\citet{Adami07a}, i.e., global scale represented in terms of the
core, intermediate, and outskirt of the cluster.
In this context, we conclude that the
\textit{shape} of the LF is very similar
in the cluster regardless of the field location over the
very wide magnitude range so far investigated ($-23<M_R<-10$).

The characteristic magnitude of the best-fit Schechter function
(\citealt{Sch76})
to Coma galaxies is $M^{*}_R\sim-20.5$ (\citealt{Mob03}).
Since most of our galaxies are much fainter than this magnitude,
we quantify our LFs with the logarithmic slope $\alpha$
which is represented,  as in \citet{AC02}, by

\begin{equation}
\label{eq:LF_a}
\alpha=-\frac{1}{0.4}\frac{\partial \log \phi}{\partial M}-1,
\end{equation}
In practice, we derive $\alpha$ by the linear regression method as
\begin{equation}
\label{eq:LF_a2}
\log \phi=-0.4(\alpha+1)M+const.
\end{equation}
not by fitting the Schechter function to the LF.

The fitting procedures are performed for
the whole range of $-19<M_R<-10$, and for the BDC and for the FDC,
separately.
We summarize the derived slopes $\alpha$ in Table \ref{tab:alphaR}.
Although slopes of the three fields differ slightly,
they are consistent within errors.
The FDC has a much steeper slope ($\alpha<-3$) than the
BDC ($\alpha\sim-1.6$).

\begin{figure}
\centering
\includegraphics[width=5cm, angle=270]{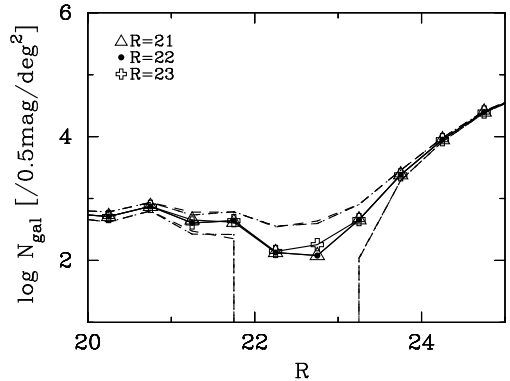}
\caption{
The  $R$-band LFs in the Coma 1 (\textit{filled circles})
when we set the star/galaxy discrimination limit to $R=21$ (\textit{open triangles}),
$R=22$ (\textit{filled circles}) and $R=23$ (\textit{open crosses}).
The dashed lines indicate errors in Poisson statistics.
All counts are scaled to an area of 1.0 deg$^2$.
\label{fig:Rlf_sep}
}
\end{figure}

We did not perform star/galaxy separation below the
limit of $R=22$ where many of our galaxies are too small to be distinguished
from stars or globular clusters (GCs).
The counts of Galactic stars fainter than this limit
in the Coma Cluster
are subtracted statistically
using the corresponding star counts in the SDF.
Figure \ref{fig:Rlf_sep} shows how the LF of Coma 1 field
changes when we adopt different limits.
The LFs for $R=21$, $R=22$ and $R=23$ are all within the errors.
We discuss the contamination from GCs for details in \S \ref{sec:gc}.

%%%%Table 3
%\clearpage
%\vspace{3cm}
\begin{table}
\begin{center}
{\scriptsize
\begin{tabular}{cccc}
\hline
         & whole             &           BDC        &  FDC  \\
Field 	&$-19<M_R<-10$	&	$-19<M_R<-13$	&	$-13<M_R<-10$	\\
\hline
Coma 1	 &$-1.77\pm0.04$	&	$-1.58\pm0.14$	&	$-3.38\pm0.28$	\\		
Coma 2	 &$-1.89\pm0.05$	&	$-1.73\pm0.22$	&	$-3.58\pm0.25$	\\
Coma 3	 &$-1.84\pm0.07$	&	$-1.54\pm0.25$	&	$-3.60\pm0.42$	\\
\hline
\end{tabular}
}
\caption[]{
List of the faint-end slope $\alpha$ in each LF for $R$-band.
}
\label{tab:alphaR}
\end{center}
\end{table}

%%%%%figure 11
%\clearpage
%%%%

%%%%figure 12
%\clearpage
%%%%

%%%%%%%%%%%%%%
\subsection{Comparison with LFs from Previous Studies}

\begin{figure}
\centering
\includegraphics[width=6cm, angle=270]{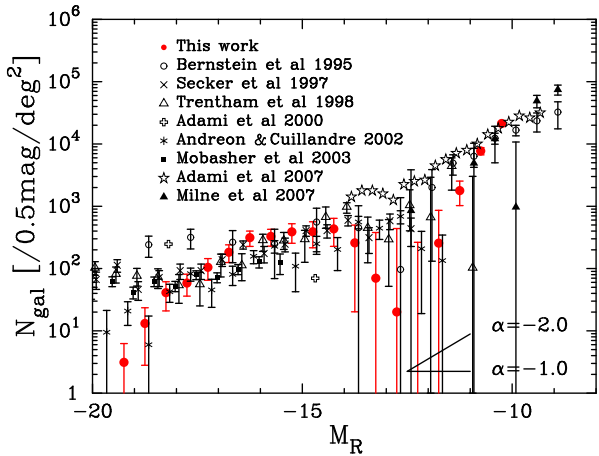}
\caption{
The $R$-band LFs of galaxies in the Coma Cluster from the literature.
The red circles indicate our LF in total area of  Coma 1, Coma 2 and Coma 3.
The ordinate shows logarithmic counts per 0.5 mag bin per deg$^2$. 
\label{fig:Rlf_pre}
}
\end{figure}

Many observational studies have been performed on the Coma Cluster LFs.
The LFs of the Coma Cluster in the $R$-band from previous studies are shown in 
Figure \ref{fig:Rlf_pre}.
Our total LF (\textit{red circles}) (Coma 1 + Coma 2 + Coma 3)
covers an unprecedentedly wide magnitude range.

The LF of our BDC agrees with most of the previous determinations down to 
$M_R\sim-13$,
and the LF of our FDC is broadly
consistent with the LFs given by \citet{Mil07}, \citet{Ber95}. 
The LFs from \citet{Mil07} and \citet{Ber95} are based on observations of a 
much smaller region in the cluster core.
Their effective areas are only $<5$\% of the field of our Coma 1 field.
Using our wide and deep imaging data, 
we confirm that the LF in the FDC of the Coma Cluster is rising steeply, 
similar to the findings of \citet{Mil07} and \citet{Ber95}, at $M_R>-13$.
The plots of  \citet{Adami07b} come from the data of their field 2 ($\sim70$ arcmin$^2$) 
and it is inconsistent with our result.

Their LF have no clear gap which distinguishes between the BDC and the FDC,
although the small dip is found at $M_R\sim-13$.
However, their LFs in the central strip region  ($\sim300$ arcmin$^2$) 
of the Coma Cluster (including the field 2),
seems to be strongly dependent on the local environments. 
Actually, their LF in the field 2 is very different from 
that in field 1, next to field 2 toward west.
The LF of their field 1 has deep dip at $M_R=-13$, 
and it has no FDC due to the contribution from GCs. 
The difference between their results and ours may be due to the difference in the size of 
observed area, i.e., a small area in the cluster core would be susceptible 
to the effect the local environment associated with the bright cD galaxy.
Our LF is consistent with most of the previous results and
we show clearly that it is composed of  well-defined two components of the BDC and the FDC.

%%%%figure 13
%\clearpage
%%%%

%%%%%%%%%%%%%
\subsection{Properties of BDC and FDC galaxies}

\begin{figure*}
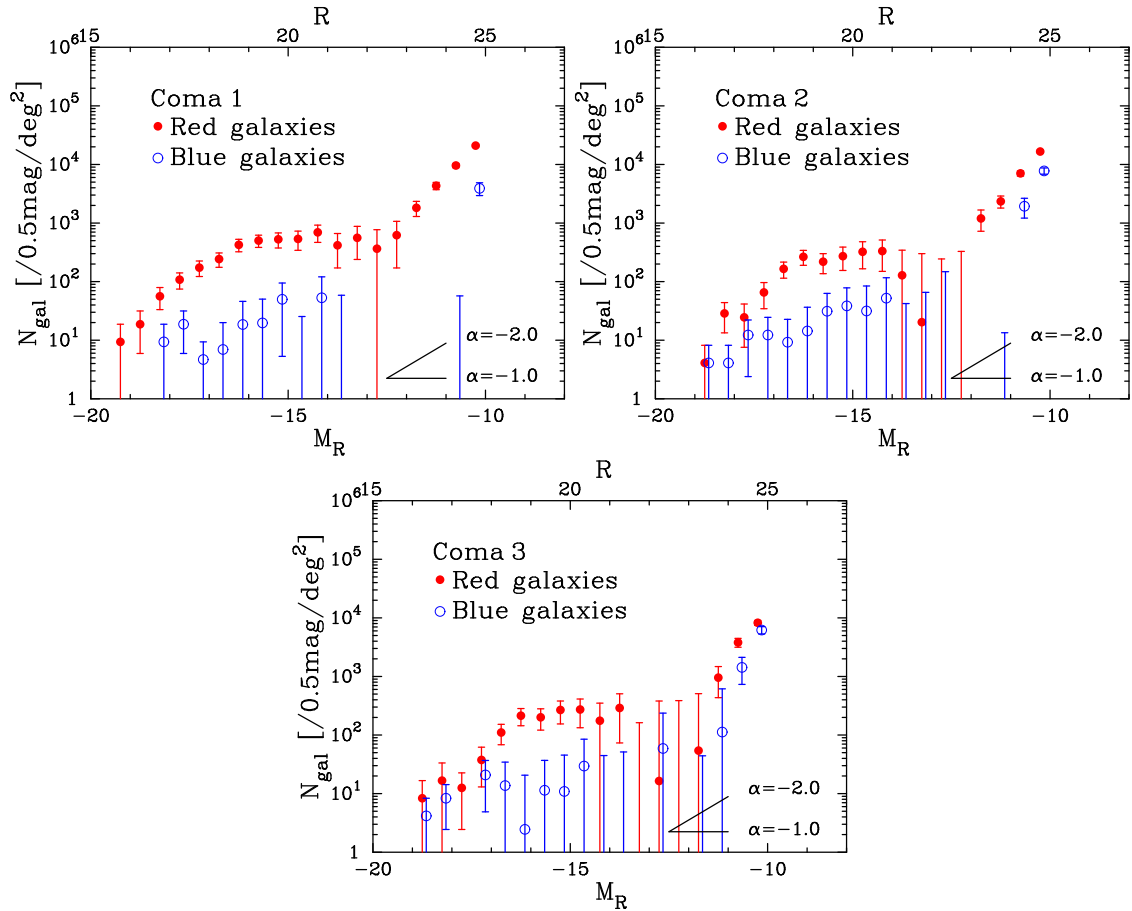

\centering
\includegraphics[width=6cm, angle=270]{f14_coma1.ps}
\includegraphics[width=6cm, angle=270]{f14_coma2.ps}
\includegraphics[width=6cm, angle=270]{f14_coma3.ps}
\caption{
The red (\textit{filled red points}) and blue galaxy (\textit{open blue points}) LFs 
of $R$-band in each field of the Coma Cluster.
\label{fig:Rlf_redblue}
}
\end{figure*}

First, we investigate the color of dwarf galaxies.
Figure  \ref{fig:Rlf_redblue} shows the LF of red and blue
galaxies defined in \S \ref{sec: color_classification} in the
three fields. It is clear from Figure  \ref{fig:Rlf_redblue}
that red galaxies are the dominant galaxy population in
the BDC of all the three fields. The small fraction of blue
galaxies in the BDC ($<10$\%) is nearly the same in the three fields.
On the other hand, the FDC of the Coma 1 is dominated by red
galaxies, while those of the Coma 2 and 3 contain some
blue galaxies.
The fraction of blue galaxies at the faintest magnitude
bin ($-11<M_R<-10$) is $\sim10$\% in the Coma 1 field
while it is $\sim40$\% in the Coma 3 field. 

\begin{figure*}
\centering
\includegraphics[width=5cm, angle=0, trim=25 20 25 10]{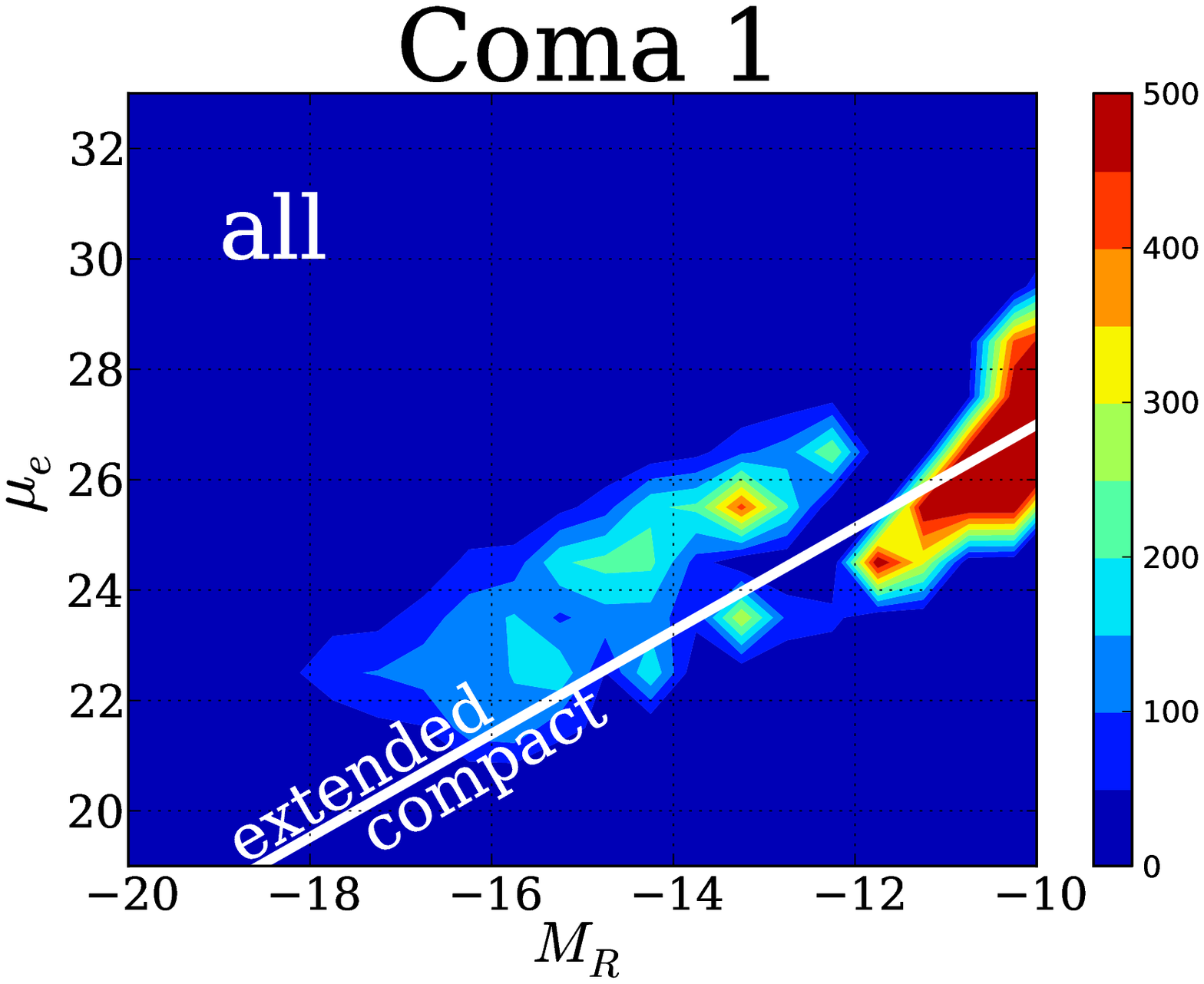}
\includegraphics[width=5cm, angle=0, trim=25 20 25 10]{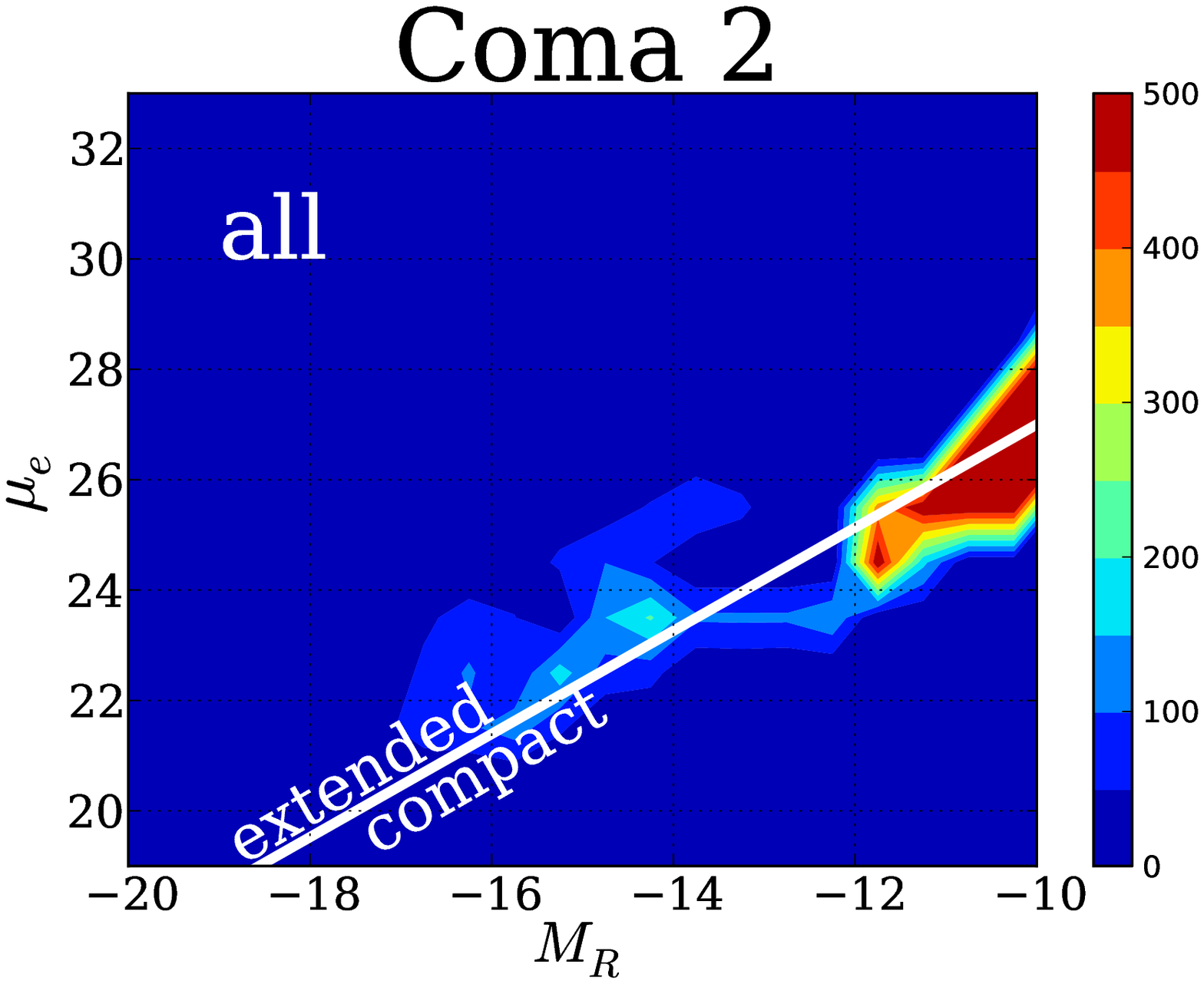}
\includegraphics[width=5cm, angle=0, trim=25 20 25 10]{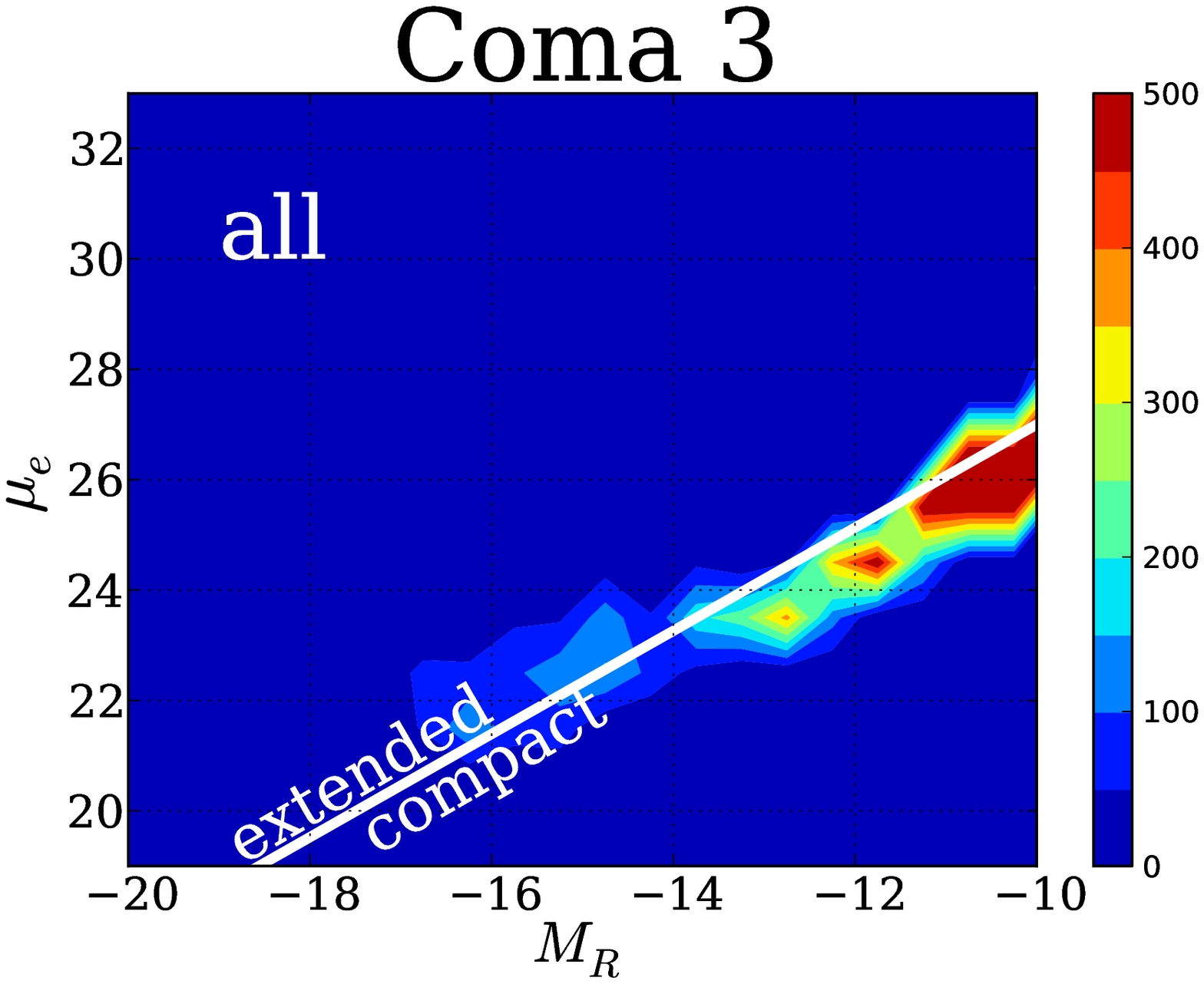}
\includegraphics[width=5cm, angle=0, trim=25 20 25 20]{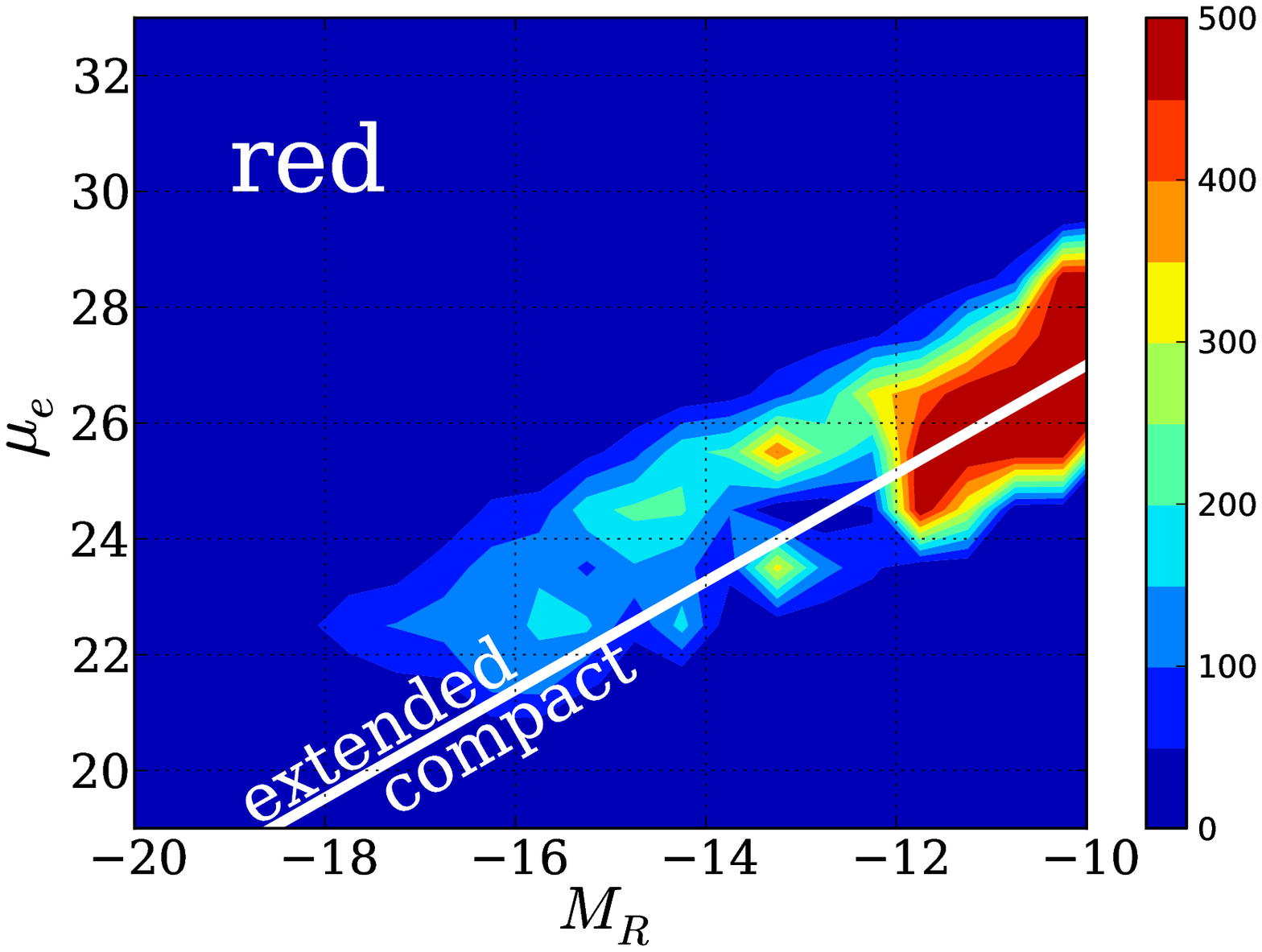}
\includegraphics[width=5cm, angle=0, trim=25 20 25 20]{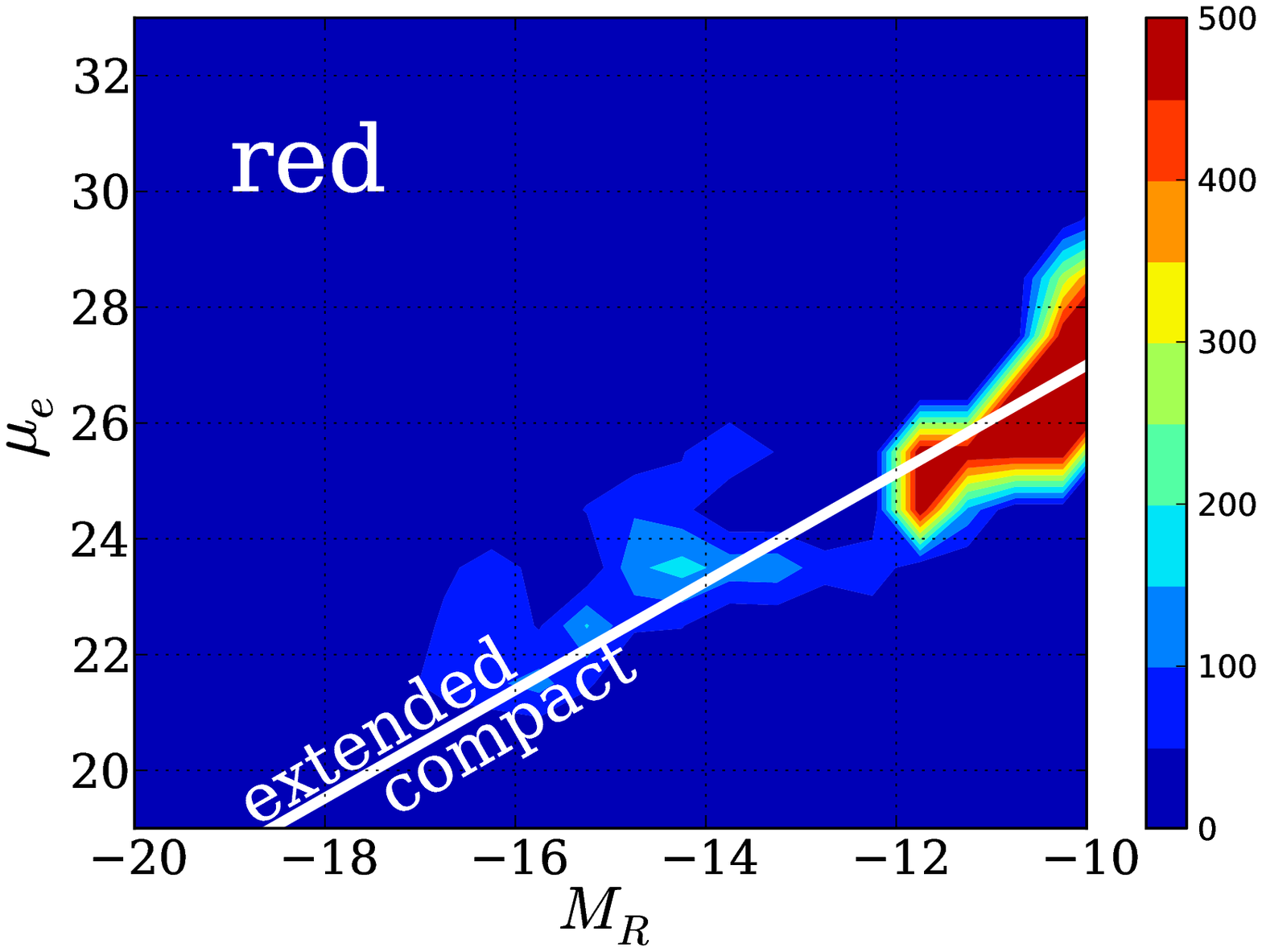}
\includegraphics[width=5cm, angle=0, trim=25 20 25 20]{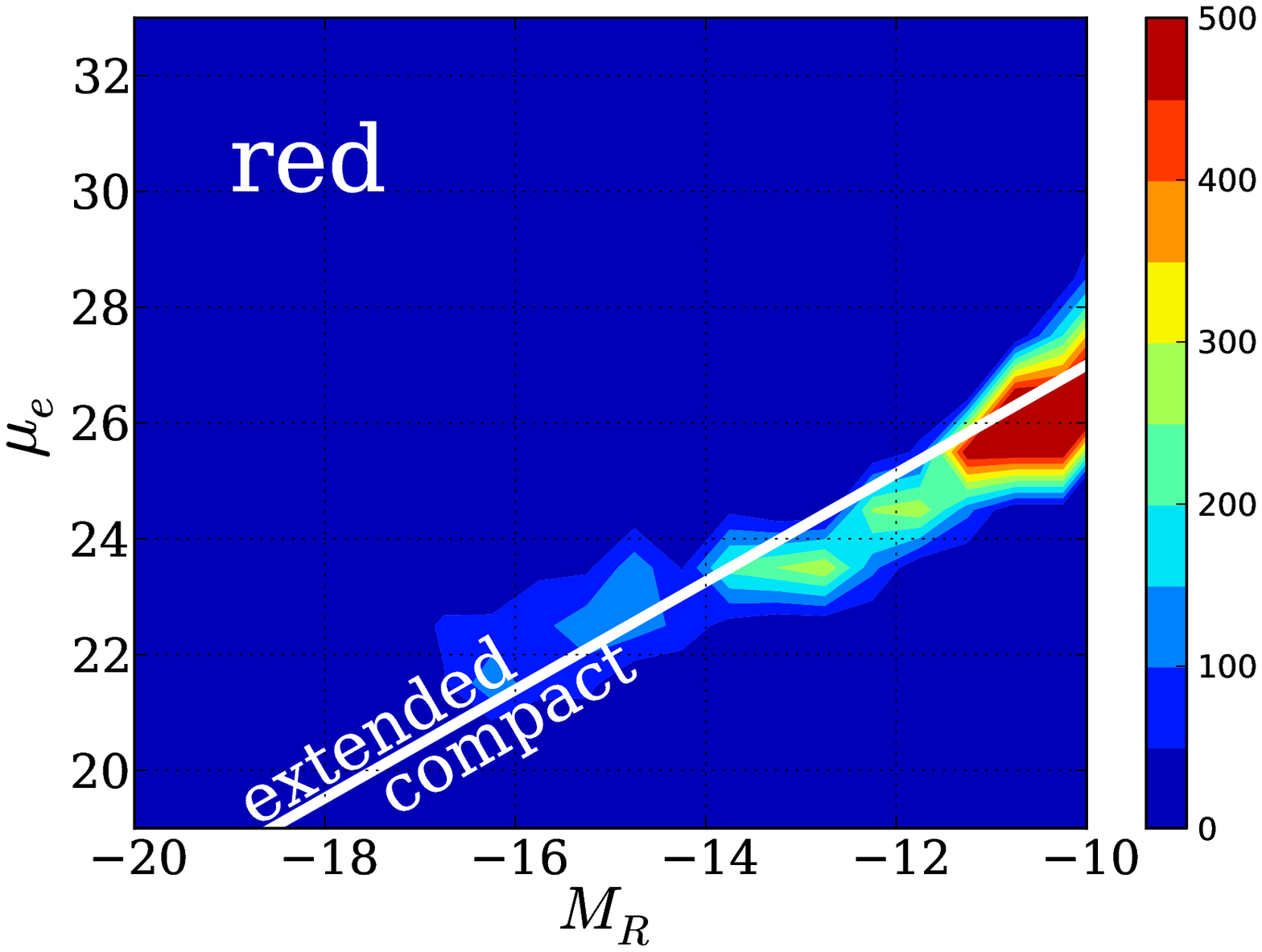}
\includegraphics[width=5cm, angle=0, trim=25 20 25 20]{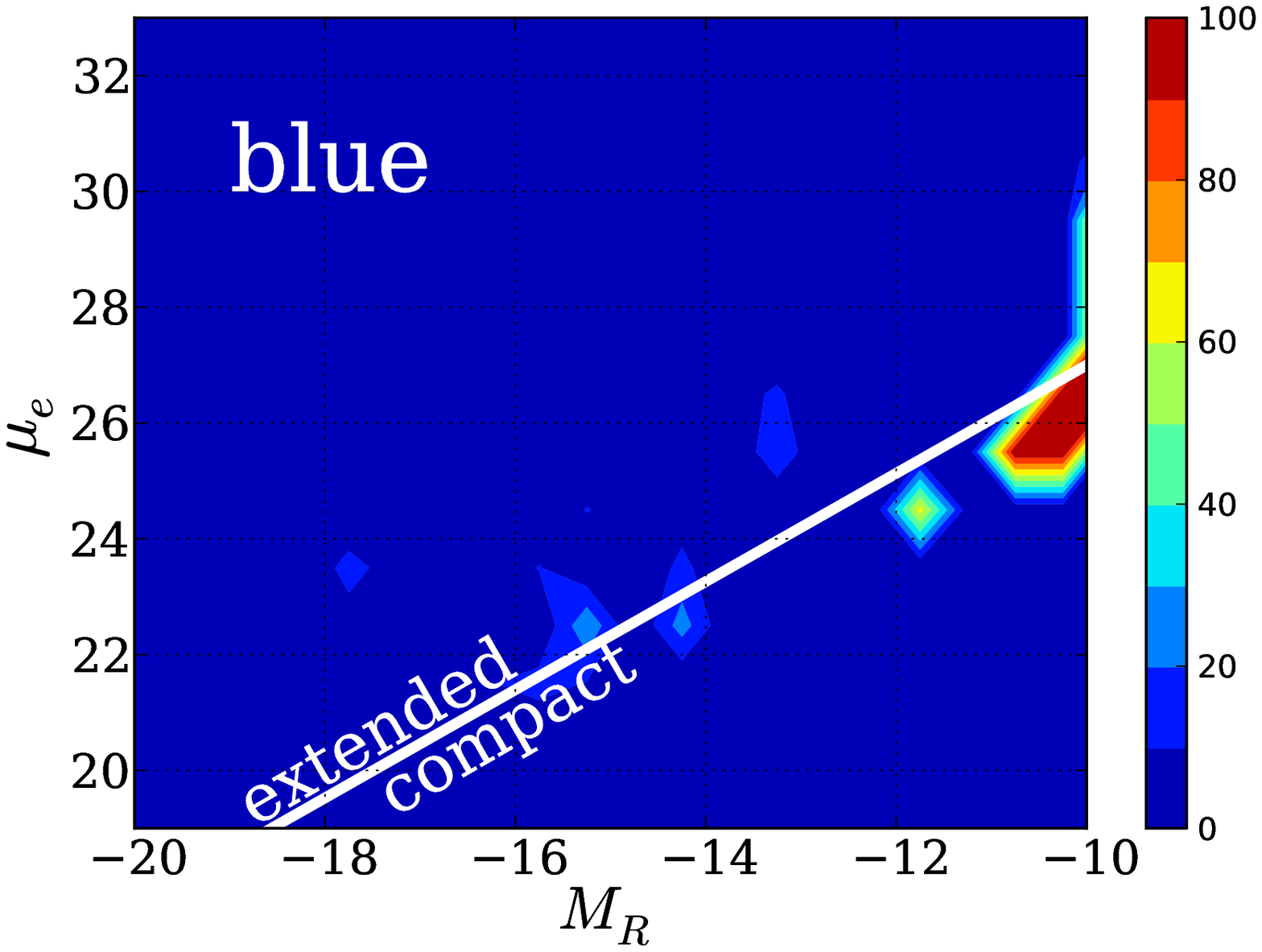}
\includegraphics[width=5cm, angle=0, trim=25 20 25 20]{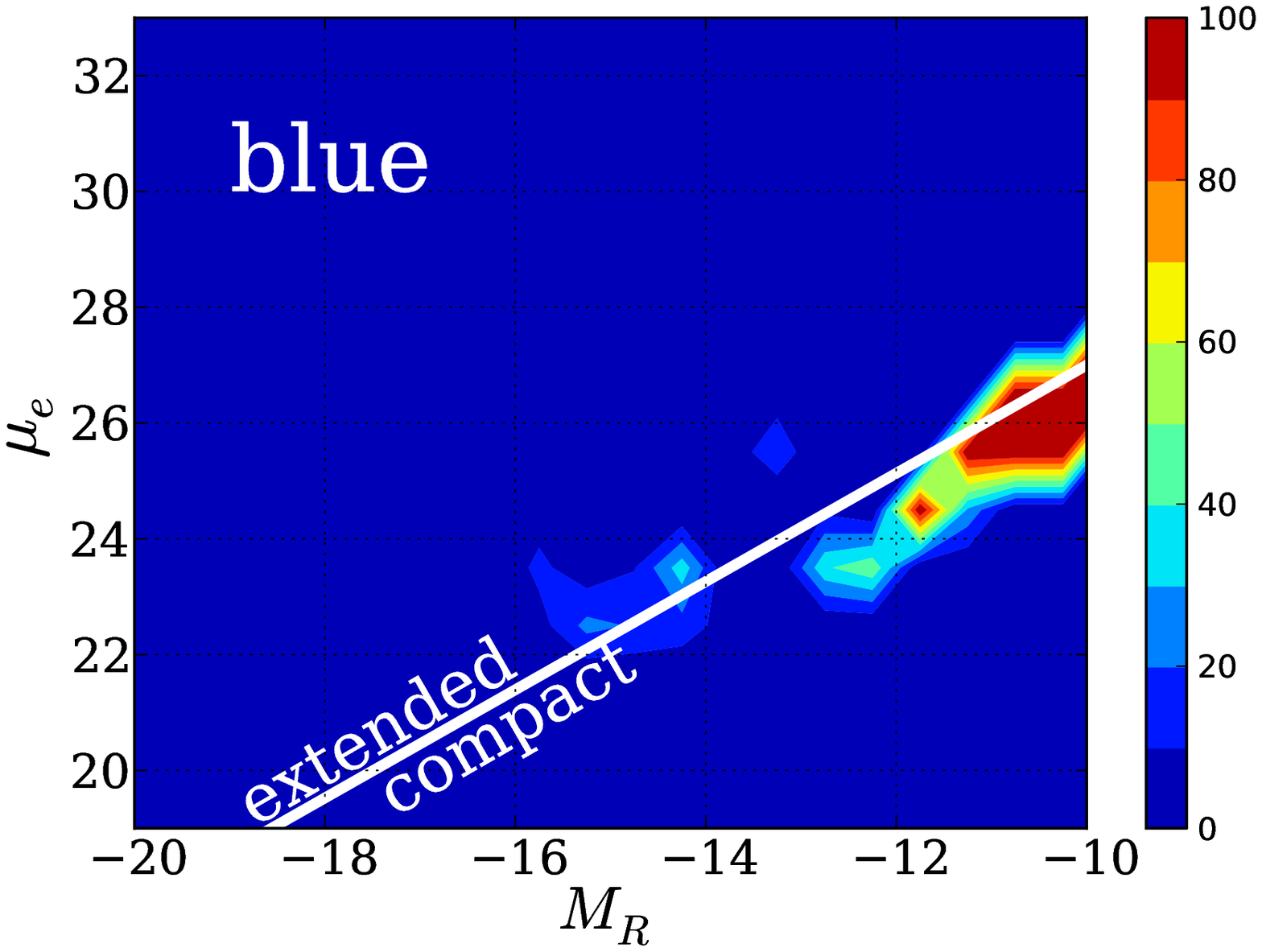}
\includegraphics[width=5cm, angle=0, trim=25 20 25 20]{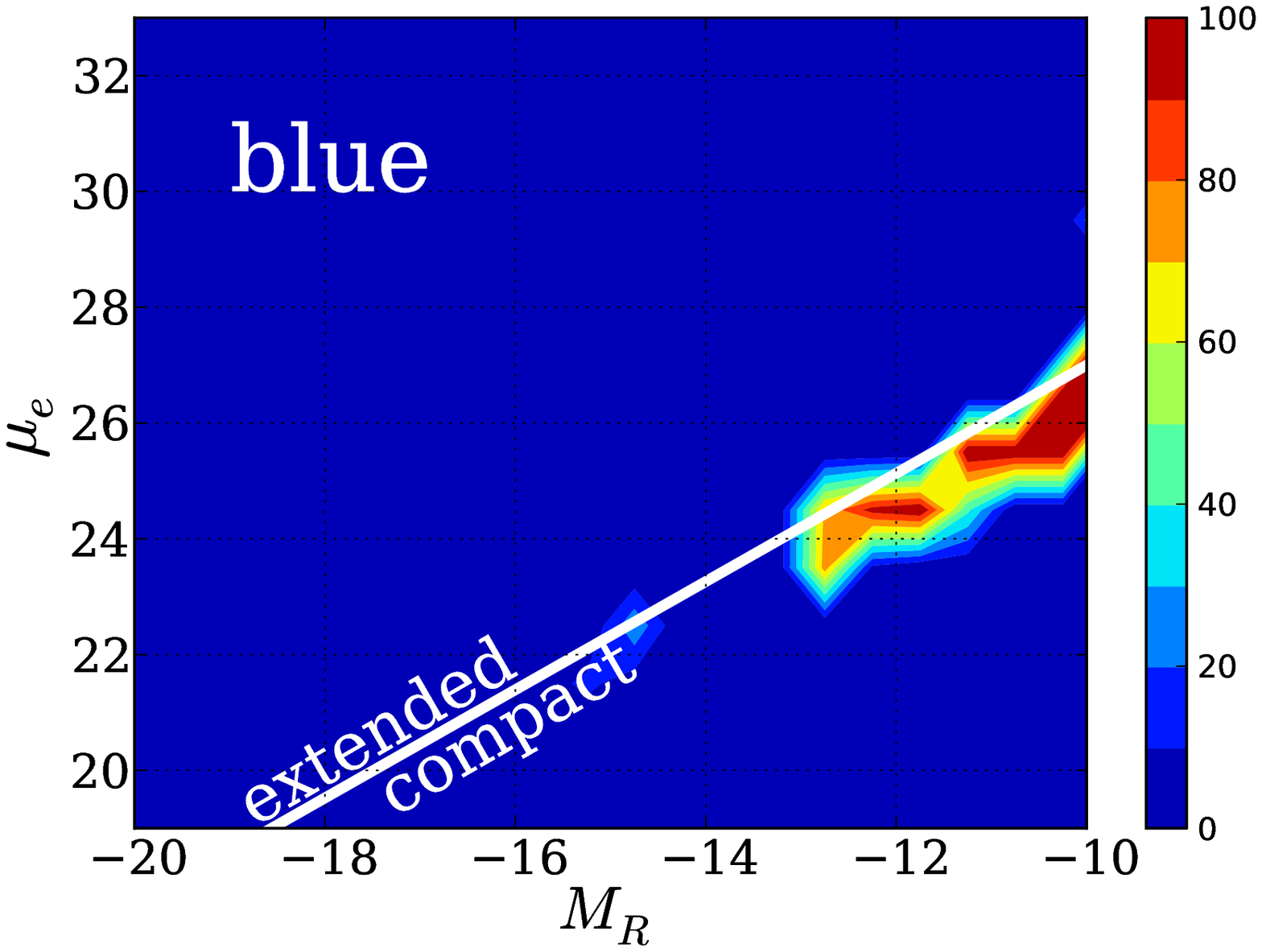}
\caption{
The color-coded number density contours of estimated member galaxies
(all galaxies: \textit{top row}, red galaxies: \textit{middle row}, blue galaxies: \textit{bottom row})
in the $R$ absolute magnitude versus effective surface brightness plane.
Coma 1, Coma 2 and Coma 3 fields are in left, middle and right columns, respectively.
The statistical background subtraction is carried out here.
The number density in each mesh is normalized to 1.0 deg$^2$ 
and corrected for the detection completeness.
The white solid line indicates the discrimination between the extended and the compact objects.
\label{fig:R_sb}
}
\end{figure*}

Next, we investigate the relationship between the magnitude and
the surface brightness. Figure \ref{fig:R_sb} shows the color-coded
density contour map in the magnitude versus surface brightness plane.
The left, middle, and right columns are for the Coma 1, Coma 2
and Coma 3 fields, respectively, while the top, middle and bottom
lows are for all, red and blue galaxies, respectively.
The white line in Figure \ref{fig:R_sb} is the line of star-like
objects in Figure \ref{fig:Rmu} shifted upward by 1.0 mag arcsec$^{-2}$,
 $\mu_e = 0.93{M_R}+36.2$. This line is used here to discriminate
between extended (low surface brightness) objects
and compact (high surface brightness) objects.
We find both extended and compact objects in the Coma 1 field, 
but very few extended objects are seen in the Coma 2 and Coma 3 fields.

Deep spectroscopic studies of faint galaxies in the Coma Cluster (e.g., \citealt{Adami09,Chib10})
help us to verify our photometric results derived from 
the statistical background subtraction. 
\citet{Adami09} presented the faint galaxies ($-14<M_R<-12$) which
are spectroscopically confirmed cluster members,  follow the CMR of 
giant member galaxies and their surface brightness 
spans approximately $2$ mag arcsec$^{-2}$ at $M_R=-13$.
In addition, they found that the compact objects are a minor component which
represent at most 5\% of the member galaxies.
This is consistent with our results of Coma 1 of Figure \ref{fig:R_sb}.
The population of our compact objects is much smaller than that of extended ones
down to $M_R=-12$.
Also \citet{Chib10} reported the spectroscopic results of dwarf galaxies ($-16<M_R<-11$)
in the core region of the Coma 1.
They examined the galaxy membership in terms of their properties such as 
morphology, surface brightness, size, and color using 140 dwarf candidates.
Their magnitude limit is $M_R=-13$ for the  low surface brightness objects and
$M_R=-11$ for the high surface brightness objects, 
while our limit is $M_R=-10$ for both of the low and high surface brightness objects
although we have no spectroscopic confirmation.  
They obtained the confirmed members of about 50 low surface brightness galaxies 
(LSBGs) with magnitudes in the range $-16<M_R<-13$,
and about 20 ultra compact dwarfs (UCDs) in the range $-13<M_R<-11$.
Their results show that the difference of  the surface brightness between LSBGs
and UCDs is about $2$ mag arcsec$^{-2}$ at $M_R=-13$
(see Figure 13 of \citealt{Chib10}). 
We also find that the difference of the surface brightness between our extended galaxies 
(objects above the white line in Figure \ref{fig:R_sb})
and compact galaxies (objects below the white line in Figure \ref{fig:R_sb})
at  $M_R=-13$ in the Coma 1 field is about 2 magnitude,
consistent with their value.
Additionally, the magnitude distributions of their UCDs are also consistent with 
our compact galaxies.

Our wide field survey revealed the following.
Most of BDC galaxies are extended and red.
The number of such galaxies increases toward the cluster center.
Very few blue extended galaxies are found in all the three fields.
On the other hand, FDC galaxies are predominantly compact
with sizes (FWHM) comparable to the seeing size,
corresponding to $\sim0.45$ kpc at the distance of the Coma Cluster.
Although most of them are red, some are blue.
Especially, a lot of blue compact galaxies exist in the FDC
of Coma 3 compared to Coma 1.
In contrast, almost no extended galaxies is found in the
FDC of Coma 3. 
The extended galaxies of FDC seem to exist only in
the dense regions such as the cluster center.

%%%%figure 14
%\clearpage
%%%%%

%%%%figure 15
%\clearpage
%%%%

%%%%%%%%%%%%%%%%%%%%%%%%%%%%%%%%%%%%%%%%%%%%%
%%%%%%%%%%%%%%%%%%%%%%%%%%%%%%%%%%%%%%%%%%%%%
%%%%%%%%%%%%%%%%%%%%%%%%%%%%%%%%%%%%%%%%%%%%%
%%%%%%%%%%%%%%%%%%%%%%%%%%%%%%%%%%%%%%%%%%%%%
%%%%%%%%%%%%%%%%%%%%%%%%%%%%%%%%%%%%%%%%%%%%%

\section{DISCUSSIONS}

First, we compare our total LF of the Coma Cluster with those of nearby
clusters taken from the literature.
The slope of the faint part of LF, $\alpha$, for nearby clusters
are summarized in Table \ref{tab:clusters}.
Note that we re-computed the values of $\alpha$ for previous studies by
fitting Equation (\ref{eq:LF_a2}) to the data values given
in respective papers, assuming that
$H_0=70$ km s$^{-1}$ Mpc$^{-1}$, $\Omega_m=0.3$, and
$\Omega_\Lambda=0.7$.
The slopes were derived by fitting in two different magnitude
ranges, $-15.0\leq M\leq-12.5$ and $-12.5\leq M\leq-10.0$
corresponding to the BDC and the FDC, respectively.
The two fitting ranges are fixed regardless of the passbands used.
We found that the LF of the Centaurus Cluster is similar to
that of the Coma Cluster showing the clear sign of BDC and FDC
and similar values of the slopes.

Next, we examine the correlation between the faint-end slope
($\alpha$) and other cluster properties,
i.e., the redshift ($z$), the Bautz--Morgan type,  the velocity
dispersion ($\sigma$),
the X-ray temperature ($T_X$) and luminosity ($L_X$).
The correlation coefficients between $\alpha$ and the
cluster parameters
are listed in Table \ref{tab:cl_cc}.
We evaluate the p-value, based on the two-tailed test for
the Pearson product-moment correlation coefficient.
No significant correlation is confirmed with $z$, BM type,
$\sigma$, or $L_X$ for either of the two components,
except for a possibly significant correlation with $T_X$
for the FDC ($-12.5\leq M\leq-10.0$). 
The slope tends to become steeper for clusters
with higher temperatures.
The field LF has a much flatter slope ($\alpha\sim-1.3$) than
the slope of the cluster LFs (e.g., \citealt{Tre05,Blan05,Liu08}).
The X-ray temperature is related to the cluster
mass (\citealt{RB02,Sta06}).

Our comparison and correlation analysis mentioned above have
led us to the conjecture that the LF in the cluster is determined
by the intrinsic properties and the evolutionary stage.
Slopes of the BDC and FDC might be related to the cluster global mass. 
In massive clusters, there is a tendency for galaxies
in the BDC to have a flatter LF and for galaxies in the FDC to
have a steep one.
Such features of the LFs of high-mass clusters may reflect
some physical processes that influenced the formation and evolution
of dwarf galaxies, such as increased intracluster medium pressure, tidal
interaction with crossing galaxies, and mergers.

The cold dark matter scenario assumes that dwarf galaxies are formed
very early in the Universe.  Based on cosmological simulations,
\citet{RG05} and \citet{BR09} showed that dwarf galaxies
forming at $z>6$ were faint dwarf spheroidal galaxies in the
present day universe. 
Their "primordial scenario" has dwarf galaxies
starting close to their current stellar mass of about
$10^{3-6}$ M$_{\bigodot}$. 
Some of these galaxies might be included in the FDC of the
Coma Cluster.

We found that colors of dwarf
galaxies in the Coma Cluster seem to be related to the environments.
A large number of red galaxies are found in the center of the Coma
Cluster while they get fewer towards outside.
The "tidal stripping scenario" (\citealt{May01,BD06})
predicts that low-luminosity galaxies would be
formed from more massive gas-rich galaxies.
In the center of massive clusters, many tidally disrupted debris
of such gas-rich galaxies are
produced (\citealt{Hen08}, \citealt{Bou08})
and their gas is effectively removed by ram pressure stripping
(e.g., \citealt{GG72,FS80,Take84,Mar03,Goto05,MB00}).
This effect may explain the existence of many red dwarf galaxies
in the center of massive clusters.

We also find a large number of compact galaxies, which have comparable
sizes to PSF, in all the three observed fields.
Recent studies suggest the existence of a number of UCD galaxies
in the core of the Coma Cluster (\citealt{Chib10, Mad10}).
It is natural to assume that the compact population represented
in the FDC of the LFs contains the UCDs.
Several studies suggest that the UCDs are the remnant
nuclei of dwarf elliptical galaxies that have lost their
outer parts due to the
tidal disruptions during passages close to the central cluster galaxy,
in a process called "galaxy threshing" (e.g., \citealt{Bekki03,Mie04,Gregg09}).
Meanwhile the formation of UCDs is linked to the merger of
young massive star clusters formed during galaxy mergers
(e.g., \citealt{FK02, Mara04}).
Thus, major galaxy mergers form a large number of compact stellar
objects that are
progenitors of GCs and UCDs (\citealt{Bou08}).
\citet{Mad10} found that UCD candidates in the Coma Cluster
are not only red but also blue.
If the UCDs originate from the interactions of galaxies,
the faint end of LFs in massive clusters like the Coma Cluster are
expected to have steep slopes
due to the strong tidal field and frequent galaxy--galaxy mergers.

We find that the outskirts contain a large fraction of blue galaxies of $-11<M_R<-10$
compared to the central region. 
\citet{Smith09} found that the passive dwarf galaxies are younger in the outskirt than those in
the core of Coma Cluster, and supported the scenario in
which many these galaxies are the quenched remnants of infalling late-type galaxies.
If late-type dwarf galaxies in clusters were star-forming galaxies infalling from outside, 
the steep slopes in the blue galaxy LFs and the large number of blue dwarfs 
in the outskirts of the cluster would be explained.

In summary, the faint part of LF in the Coma Cluster seems to consist of some populations of galaxies 
whose origins are different. 
We suggest that these red galaxies contain primordial dwarf galaxies,  
UCDs of the tidal origin and remnants of disrupted dwarf galaxies,
while blue dwarf galaxies contain some blue UCDs
and infalling late-type dwarf galaxies from the outer fields.
In order to understand their origin, the spectroscopic information such as their
dynamics and metallicity is essential.
Further investigations, we require the spectroscopic observations of very faint galaxies
down to $M=-10$ in various regions of the Coma Cluster to identify member galaxies.

%%%%Table 4
%\clearpage
\begin{table*}
%\begin{landscape}
\begin{center}
{\tiny
\begin{tabular}{cccccccccc}
\hline
Cluster	   &z	   &BM	   &$\sigma$      &$L_X$	 &$T_X$	&Band	&$\alpha$	&$\alpha$	 &Reference	\\
	   &	   &	   &(km s$^{-1})$    &($10^{44}$erg s$^{-1}$) &(keV)&	&($-15\leq M\leq-12.5$)&($-12.5\leq M\leq-10.0$)&	\\	
\hline
Virgo	   &0.0036$^\mathrm{a}$ &III   & 643$^\mathrm{e}$&0.418$^\mathrm{g}$  &2.28$^\mathrm{g}$&$B$&$-1.61\pm0.07$ &$-0.56\pm0.10$  &\citealt{Sab03} \\
     	   &0.0036              &III   & 643	         &0.418		    &2.28 	      &$B$&$\sim-1.5	$ &$\sim-1	$  &\citealt{TH02} \\
     	   &0.0036              &III   & 643	         &0.418		    &2.28 	      &$R$&$-2.06\pm0.06$ &$-2.07\pm0.24$  &\citealt{Phi98} \\
Fornax	   &0.0046$^\mathrm{b}$ &I     & 374$^\mathrm{f}$&0.00119$^\mathrm{h}$&1.56$^\mathrm{k}$&$B$&$\sim-2	$ &$\sim-2      $  &\citealt{Kamb00}\\
    	   &0.0046              &I     & 374	         &0.00119		    &1.56 	      &$V$&$-1.38\pm0.06$ &$-1.26\pm0.01$  &\citealt{Hilk03}\\
Centaurus  &0.0114$^\mathrm{c}$ &I-II  & 863$^\mathrm{c}$&1.378$^\mathrm{h}$  &3.69$^\mathrm{k}$&$V$&$-0.69\pm0.13$ &$-3.73\pm0.03$  &\citealt{Chib06}\\
Hydra	   &0.0126$^\mathrm{c}$ &III   & 647$^\mathrm{c}$&0.569$^\mathrm{h}$  &3.15$^\mathrm{k}$&$B$&$-1.88\pm0.04$ &$-1.38\pm0.08$  &\citealt{Yama07}\\
           &0.0126              &III   & 647	         &0.569	            &3.15 	      &$R$&$-1.48\pm0.02$ &$-1.64\pm0.03$  &\citealt{Yama07}\\
Perseus	   &0.0179$^\mathrm{c}$ &II-III&1324$^\mathrm{c}$&15.341$^\mathrm{i}$ &6.42$^\mathrm{k}$&$B$&$-1.81\pm0.04$ &$-1.00\pm0.02$  &\citealt{Con02} \\
           &0.0179              &II-III&1324	         &15.341              &6.42 	      &$B$&$-1.53\pm0.09$ &$-	        $  &\citealt{PC08}  \\
Coma	   &0.0231$^\mathrm{c}$ &II    &1008$^\mathrm{c}$&7.767$^\mathrm{i}$  &8.25$^\mathrm{l}$&$B$&$-1.28\pm0.03$ &$-3.50\pm0.18$  & This work      \\
           &0.0231              &II    &1008	         &7.767		    &8.25 	      &$R$&$-0.31\pm0.12$ &$-3.97\pm0.05$  & This work      \\
A2199	   &0.0302$^\mathrm{d}$ &I     & 733$^\mathrm{c}$&4.079$^\mathrm{j}$  &3.99$^\mathrm{j}$&$B$&$-2.24\pm0.09$ &$-1.65\pm0.51$  &\citealt{DeP95} \\
\hline
\end{tabular}
}
\caption[]{
List of parameters of nearby clusters.
Note that the magnitude is converted into the AB system adopting
$H_0=70$ km s$^{-1}$ Mpc$^{-1}$, $\Omega_m=0.3$, and $\Omega_\Lambda=0.7$.
The $L_X$ and $T_X$ from the X-Rays Clusters Database (BAX) 
are set to $H_0=50$ km s$^{-1}$ Mpc$^{-1}$, $\Omega_m=1.0$.
}
\label{tab:clusters}
\end{center}
{\footnotesize
z: redshift \\
--- $^\mathrm{a}$\citet{Ebe98}, $^\mathrm{b}$\citet{Abel89}, 
$^\mathrm{c}$\citet{SR99}, $^\mathrm{d}$\citet{OH01} \\
BM: Bautz-Morgan type (NED) \\
$\sigma$: velocity dispersion \\
--- $^\mathrm{e}$\citet{Zab90}, $^\mathrm{f}$\citet{Drink01} \\
$L_X$: X-ray luminosity in the 0.1-0.2 keV band \\
--- $^\mathrm{g}$\citet{Matsu00}, $^\mathrm{h}$\citet{Boh04}, $^\mathrm{i}$\citet{RB02},
$^\mathrm{j}$\citet{Vik09}  \\
$T_X$: X-ray temperature \\
--- $^\mathrm{k}$\citet{Ike02}, $^\mathrm{l}$\citet{Arn01} \\
$\alpha$: faint-end slope of LF
}
%\end{landscape}
\end{table*}

%%%%Table 5
%\clearpage
%\vspace{4.5cm}
\begin{table*}
\begin{center}
\begin{tabular}{ccccc}
\hline
parameter	& \multicolumn{2}{c}{$\alpha$($-15\leq M\leq-12.5$)} & \multicolumn{2}{c}{$\alpha$($-12.5\leq M\leq-10$)} \\ 
	        & correlation coefficient & p-value &  correlation coefficient & p-value \\
		
\hline
z		&$0.14$   &$0.64$   &$-0.43$   &$0.16$ \\
BM		&$-0.08$  &$0.79$   &$0.33$    &$0.27$ \\
$\sigma$	&$0.32$   &$0.29$   &$-0.34$   &$0.26$ \\
$log(L_X)$		&$0.23$   &$0.45$   &$-0.31$   &$0.31$ \\
$T_X$		&$0.49$   &$0.09$   &$-0.61$   &$\bf 0.03$ \\ 
\hline
\end{tabular}
\caption{
List of correlation coefficients between $\alpha$ and the parameter.
Note that the p-value under 0.05 is considered significant. 
}
\label{tab:cl_cc}
\end{center}
\end{table*}

%%%%%%%%%%%%%%%%%%%%%%%%%%%%%%%%%%%%%%%%%%%%
%%%%%%%%%%%%%%%%%%%%%%%%%%%%%%%%%%%%%%%%%%%%
%%%%%%%%%%%%%%%%%%%%%%%%%%%%%%%%%%%%%%%%%%%%
%%%%%%%%%%%%%%%%%%%%%%%%%%%%%%%%%%%%%%%%%%%%
\section{CONCLUSIONS}

We construct the galaxy luminosity functions (LFs) for the Coma Cluster, the richest of
the nearby galaxy clusters. 
Deep and wide images of the Coma Cluster are obtained with the Suprime-Cam mounted 
on the Subaru Telescope.
The resultant LFs cover the range of $-19<M_R<-10$. 
Contamination from background galaxies is subtracted statistically using 
the number counts of galaxies in a blank field, the Subaru Deep Field.

Our main results are summarized as follows:
 
\begin{enumerate}
\item
We derive LFs in the cluster center region (Coma 1 field), 
the NGC 4839 group (Coma 2 field) and the outskirts (Coma 3 field).
No significant differences in the shape of the LFs are found
in the different fields of the Coma Cluster.

\item
The faint part of LF in the Coma Cluster contains two distinct components; 
the bright dwarf component (BDC; $-19<M_R<-13$) which has a relatively flat LF slope and
the faint dwarf component (FDC; $-13<M_R<-10$) which has a steep slope.

\item
Most of BDC galaxies are extended and red.
The number of such galaxies increases toward the cluster center.
Very few extended and blue galaxies are found in the BDC of
all the three fields.
The fraction of blue (both extended and compact) galaxies in
the BDC ($<10$\%) is nearly the same in all the three fields.

\item
 On the other hand, FDC galaxies are predominantly compact
with sizes comparable to the seeing size,
FWHM corresponding to $\sim0.45$ kpc at the distance of the Coma Cluster.
Most of them are red, but some are blue.
Especially, a lot of blue compact galaxies exist in the FDC
of Coma 3 compared to Coma 1.
In contrast, almost no extended galaxies is found in the
FDC of Coma 3. The extended galaxies of FDC seem to exist only in
the dense regions such as the cluster center.

\end{enumerate}

%%%%%%%%%%%%%%%%%%%%%%%%%%%%%%%%%%%%%%%%%%%%%
%%%%%%%%%%%%%%%%%%%%%%%%%%%%%%%%%%%%%%%%%%%%%
%%%%%%%%%%%%%%%%%%%%%%%%%%%%%%%%%%%%%%%%%%%%%
%%%%%%%%%%%%%%%%%%%%%%%%%%%%%%%%%%%%%%%%%%%%%
%%%%%%%%%%%%%%%%%%%%%%%%%%%%%%%%%%%%%%%%%%%%%

%%%%%%%%%%%%%%%%%%%%%%%%%%%%%%%%%%%%%%%%%%%%%%%%%%%%%%
\acknowledgments

This study is based on the PhD thesis of the first author
which was accepted in 2010
at the Graduate University for Advanced Studies (Sokendai).
We are grateful to the staff of the Subaru Telescope for their help.
We thank Margaret Milne for providing her Coma LF data, 
Nobuo Arimoto and Tadayuki Kodama for discussions.
We would like to thank anonymous referee for his/her helpful suggestions.
This research has made use of the Subaru Deep Field (SDF) archive,
the NASA/IPAC Extragalactic Database (NED),
the X-Rays Clusters Database (BAX) and the Digitized Sky Surveys (DSS).
Funding for the SDSS and SDSS-II has been provided by the Alfred P. Sloan Foundation, 
the Participating Institutions, the National Science Foundation, the U.S. Department of Energy, 
the National Aeronautics and Space Administration, the Japanese Monbukagakusho,
the Max Planck Society, and the Higher Education Funding Council for England. 
The SDSS Web Site is http://www.sdss.org/.
Data analysis were in part carried out on common use data analysis computer system 
at the Astronomy Data Center, ADC, of the National Astronomical Observatory of Japan.

%
%
%
%
%

%%%%%%%%%%%%%%%%%%%%%%%%%%%%%%%%%%%%%%%%%%%%%
%%%%%%%%%%%%%%%%%%%%%%%%%%%%%%%%%%%%%%%%%%%%%
%%%%%%%%%%%%%%%%%%%%%%%%%%%%%%%%%%%%%%%%%%%%%
%%%%%%%%%%%%%%%%%%%%%%%%%%%%%%%%%%%%%%%%%%%%%
%%%%%%%%%%%%%%%%%%%%%%%%%%%%%%%%%%%%%%%%%%%%%

%
%
%

%
%
%

%%%%%%%%%%%%%%%%%%%%%%%%%%%%%%%%
%%%%%%%%%%%%%%%%%%%%%%%%%%%%%%%%
\end{document}